\documentclass[12pt]{article}
\usepackage{epsfig,amssymb}
\usepackage{latexsym}

\newcommand{\bq}{\begin{equation}}
\newcommand{\eq}{\end{equation}}
\newcommand{\bqa}{\begin{eqnarray}}
\newcommand{\eqa}{\end{eqnarray}}
\newcommand{\ben}{\begin{enumerate}}
\newcommand{\een}{\end{enumerate}}
\newcommand{\bc}{\begin{center}}
\newcommand{\ec}{\end{center}}
\newcommand{\bqb}{\begin{eqnarray*}}
\newcommand{\eqb}{\end{eqnarray*}}

\def\gsim{\gtrsim}

\def\pr#1#2#3{ Phys. Rev. ${\bf{#1}}$, #2 (#3)}
\def\prl#1#2#3{ Phys. Rev. Lett. ${\bf{#1}}$, #2 (#3)}

\def\np#1#2#3{ Nucl. Phys. ${\bf{#1}}$, #2 (#3)}
\def\zp#1#2#3{ Z. f. Phys. ${\bf{#1}}$, #2 (#3)}
\def\jhep#1#2#3{ JHEP ${\bf{#1}}$, #2 (#3)}
\def\epj#1#2#3{ Eur. Phys. J. ${\bf{#1}}$, #2 (#3)}
\def\ijmp#1#2#3{ Int. J. Mod. Phys. ${\bf{#1}}$, #2 (#3)}

\def\fortp#1#2#3{ Fortsch. Phys. ${\bf{#1}}$, #2 (#3)}

\def\aop#1#2#3{Annals of Phys. ${\bf{#1}}$, #2 (#3)}

\global\nulldelimiterspace = 0pt

\def\swsq{s^2_W}
\def\cwsq{c^2_W}

\def\mwsq{m_W^2}
\def\mw{m_W}

\def\tchi{\tilde \chi}

\begin{document}
\pagenumbering{arabic}
\thispagestyle{empty}
\def\thefootnote{\fnsymbol{footnote}}
\setcounter{footnote}{1}

\begin{flushright}
February 14,  2011\\
arXiv:1012.1114 [hep-ph]\\
Extended version\\

 \end{flushright}
\vspace{2cm}
\begin{center}
{\Large {\bf The processes   $gg  \to \tchi^+_i\tchi^-_j, ~\tchi^0_i\tchi^0_j$
at high energies.    } } \\
 \vspace{1.5cm}
{\large G.J. Gounaris$^a$, J. Layssac$^b$,
and F.M. Renard$^b$}\\
\vspace{0.2cm}
$^a$Department of Theoretical Physics, Aristotle
University of Thessaloniki,\\
Gr-54124, Thessaloniki, Greece.\\
\vspace{0.2cm}
$^b$Laboratoire de Physique Th\'{e}orique et Astroparticules,
UMR 5207\\
Universit\'{e} Montpellier II,
 F-34095 Montpellier Cedex 5.\\
\end{center}

\vspace*{1.cm}
\begin{center}
{\bf Abstract}
\end{center}
According to the helicity conservation (HCns) theorem,
the sum of the helicities should be conserved,
in  any 2-to-2 processes in MSSM with R-parity conservation,
at high energies; i.e. all amplitudes violating this  rule,
must  vanish asymptotically. The realization   of HCns     in
 $gg  \to \tchi^+_i\tchi^-_j,\tchi^0_i\tchi^0_j$ is studied,
 at  the one loop electroweak order (EW), and   simple high energy
   expressions are derived for the  non-vanishing helicity conserving (HC) amplitudes.
  These are  very similar  to  the corresponding   expressions   for
 $gg \to W^+W^-, ~ZZ,~ \gamma Z,~ \gamma\gamma $ derived before.
 Asymptotic relations among observable unpolarized cross sections for many such processes
  are then  obtained,  some of which may  hold   at LHC-type  energies.    \\

\vspace{01cm}

\def\thefootnote{\arabic{footnote}}
\setcounter{footnote}{0}
\clearpage

\section{Introduction}

\hspace{0.7cm}In  \cite{heli1, heli2}, we have established the helicity conservation (HCns)
 theorem, to all orders in  the minimal supersymmetric model (MSSM)
 with R-parity conservation.
 This theorem states that for any 2-to-2  process, the only amplitudes that can
 survive  at asymptotic energies and a fixed angle, are  the helicity conserving (HC) ones,
in which  {\it the sum of the two initial helicities
equals to the sum of the two final ones}.
  All   amplitudes  violating    this  rule, and therefore  called helicity violating (HV)
  amplitudes,   are predicted to vanish   asymptotically.

In the general all-order proof presented in \cite{heli1, heli2},
all mass dimensional parameters
were  neglected. For a 2-to-2 processes, this is a reasonable assumption for calculating
amplitudes at a kinematical region where the SUSY-breaking effects are negligible.
Under this assumption, it was then showed that all HV amplitudes vanish asymptotically.

This  general  proof  though, gives no indication on how the asymptotically dominating
HC amplitudes behave at high energies \cite{heli1, heli2}. To see  this,
detail 1loop calculations for  specific process are needed.

In such detail calculations, we have  observed that in processes involving
external gauge bosons, huge  cancelations among the various diagrams need to conspire,
in order to  establish HCns  \cite{heli1, heli2, Corfu1}. On the contrary, no such
cancelations appear, in processes where all  external particles are fermions or scalars.
In fact, it was this property that motivated  us at first, to  look at $ug\to d W^+$
and $ug\to \tilde d_L \tchi_i^+$, at the 1loop EW order \cite{ugdW, ugsdWino}.

 Concerning  the asymptotic HC amplitudes for  $ug\to d W^+$
 at the    1loop  MSSM EW order,   we note that they were  indeed found to
   depend on the magnitude  of the SUSY masses, and thereby on the SUSY breaking
   terms \cite{ugdW, ugsdWino}.
 More explicitly, the leading logarithmic corrections were found to depend mainly on the
 average scale of the SUSY masses; while  the subleading energy-independent "constants" were
 found to depend on ratios of the internal and/or external masses,
 as well as on the scattering angle \cite{ugdW, ugsdWino}.
The situation is further complicated  when  SUSY particles appear in the final state,
like e.g. in  $ug\to \tilde d_L \tchi_i^+$ \cite{ugsdWino},
where the $\tchi^+_i$ wave function  introduces additional    dependencies  on
ratios of the SUSY breaking mass-dimensional parameters.

The 1loop EW order calculations for  $ug\to dW$
and\footnote{Here $\tilde d_L$ describe a down-L-squark and $\tchi_i^+$ a chargino.}
$ug\to \tilde d_L \tchi_i^+$, were also used to derive simple asymptotic relations
between  the unpolarized cross sections for these processes \cite{ugdW, ugsdWino}.
The particularly interesting thing in these relations is
that they may  be satisfied even   at LHC type energies, if
the SUSY scale is  within the rather wide  range  of the benchmarks
\cite{SPA1, SPA2, Baer1, Baer2, Baer3}.

Subsequently we  studied, at the same 1loop EW order, the processes
 $gg\to HH'$, $VH,~VV'$,  where two gluons fuse to produce  Higgs-type scalars
or electroweak vector  particles ($V,V'=W,Z,\gamma $), in either SM or MSSM \cite{ggHH, ggVV}.
In this case,  there are  no Born contributions, neither
any    gauge contribution; and no large logarithmic contributions to the
HC amplitudes appear  asymptotically.
As a result, the asymptotic HC amplitudes were  found to be very  sensitive to the
differences between the SM and MSSM   dynamics; but quite insensitive
to the SUSY breaking mass terms.
Thus,  examples were found, involving either longitudinal \cite{ggHH},
or transverse  gauge bosons \cite{ggVV},
where  HCns is strongly violated in SM,
while  obeyed in MSSM.  This shows that  HCns is indeed
a genuine  SUSY property, drastically reducing
the number of the asymptotically non-vanishing amplitudes \cite{heli1, heli2, CM}.

The purpose of the present work is to study the helicity amplitudes for
$gg  \to \tchi^+_i\tchi^-_j, ~\tchi^0_i\tchi^0_j$, at the 1loop EW order in MSSM.
In these processes, two gluons fuse  to produce two charginos or neutralinos,
which constitute   the supersymmetric transformed of appropriate combinations
of the  final states in   $gg\to HH',VH,VV'$ studied in \cite{ggHH, ggVV}.
 This  leads to interesting relations  among  the corresponding
 processes at high energies.

Again, very simple expressions for the asymptotic  HC amplitudes for
$gg  \to \tchi^+_i\tchi^-_j, ~\tchi^0_i\tchi^0_j$ are obtained,  similar
to the corresponding $gg\to VV'$ amplitudes found in \cite{ggVV}.
The interrelations  between the two processes,
   naturally appear in  two different forms; the gauge-gaugino relations,
and the Goldstone-higgsino ones.
In all cases, they are independent of the squark masses running along the internal lines
of the contributing diagrams. In addition, fascinating SUSY identities between
the asymptotic amplitudes for
$gg \to VV'$ and $\tilde g \tilde g  \to \tchi^+_i\tchi^-_j, ~\tchi^0_i\tchi^0_j$
are noted.

Furthermore,    simple asymptotic relations
among the unpolarized cross sections for $gg \to VV'$ and
$gg  \to \tchi^+_i\tchi^-_j, ~\tchi^0_i\tchi^0_j$  are constructed, analogous to those
in \cite{ugsdWino}. It is found that some these relations may be approximately respected,
even at LHC-type  energies.

A Fortran code supplying the 1loop helicity amplitudes  for
$gg  \to \tchi^+_i\tchi^-_j, ~\tchi^0_i\tchi^0_j$, at any energy and angle in MSSM,
  is herewith released \cite{code}.

The contents of the paper are the following:
Sect.2 is devoted to the amplitude analysis of
$gg\to \tchi_i\tchi_j $, including a presentation of the numerical
code  mentioned above.
In Sect.3,  the   high energy expressions for these amplitudes
are presented, which, together with those   for $gg\to VV'$,
  are used to illustrate  the helicity conservation theorem for these processes,
  and to derive   the aforementioned asymptotic relations
    among their   unpolarized cross sections.
    The numerical results are presented in Sect. 4.
 The concluding remarks  appear in Section 5. \\

\section{The  $gg\to \tchi^0_i\tchi^0_j,~\tchi^+_i\tchi^-_j$ amplitudes}

Defining the usual kinematical variables
\bq
s=(p_g+p'_g)^2~~,~~ t=(p_g-p_{\tchi_i})^2~~,~~ u=(p_g-p_{\tchi_j})^2 ~~, \label{kin1}
\eq
in terms of the incoming and outgoing momenta, the helicity amplitudes
are  denoted as
\bq
 F(gg\to \tchi_i\tchi_j)_{\mu \mu'\tau\tau'}\equiv F_{\mu \mu'\tau\tau'}^{ij}(\sqrt{s},\theta)
  ~~,  \label{F-def}
\eq
where $s$ is the  squared  c.m. energy, and  $\theta$
is the c.m. scattering angle between $p_g$ and $p_{\tchi_i^0}$.
The indices  $\mu, \mu', \tau, \tau'$ in (\ref{F-def}) denote respectively
the  $g,g,\tchi_i, \tchi_j$ helicities, using  the standard conventions \cite{JW};
while   $i,j$ describe the mass eigenstates of
 the final\footnote{In calculating these amplitudes, positive (negative)
 energy Dirac wave functions are always used for describing
 the first (second)  outgoing particles
  $\tchi_i$ ($\tchi_j$). }  neutralinos or charginos.
 A color-factor $\delta^{ab}$, where $a,b$ are the gluon color indices,
is always  removed from the amplitude, defined so that the $i F$-phase coincides
with the phase of the
 S-matrix.  The c.m. momentum of  the  final state particles is
\bq
p={[s-(m_i+m_j)^2]^{1\over2}[s-(m_i-m_j)^2]^{1\over2}\over2\sqrt{s}} ~~. \label{pcm-final}
\eq
The unpolarized differential cross section, as well
the dimensionless $\tilde \sigma$, are defined as
\bqa
 \tilde \sigma(gg\to \tchi_i\tchi_j)  & \equiv & \frac{512 \pi}{\alpha^2 \alpha_s^2}\,
\frac{s^{3/2}}{p}\, {d\sigma (gg\to \tchi_i\tchi_j; s) \over d\cos\theta}
=\frac{\sum_{\mu\mu'\tau\tau'}|F_{\mu\mu'\tau\tau'}(gg\to \tchi_i\tchi_j)|^2}
{\alpha^2 \alpha_s^2}~~~,  \label{sigmatilde}
\eqa
where the summation is over all initial and final helicities.
For  identical  neutralinos  $\tchi^0_i\tchi^0_i$, the integrated cross sections
in the region $-1<\cos\theta <1$,  must be divided by 2.\\

Bose and Fermi statistics for the initial gluons and the final  neutralinos  constrain
the helicity amplitudes by
\bqa
 gg  & \Rightarrow &
F^{ij}_{\mu\mu'\tau\tau'}(\theta)
= (-1)^{\tau-\tau'} F^{ij}_{\mu' \mu  \tau\tau'}(\pi-\theta)~~, \label{gg-Bose} \\
  \tchi_i^0\tchi_j^0   & \Rightarrow &
F^{ij}_{\mu\mu'\tau\tau'}(\theta)= F^{ji}_{\mu \mu'\tau'\tau}(\pi-\theta)~~.
\label{chi00-Fermi}
\eqa
In addition, when neglecting  the  CP violating contribution to MSSM, we  also obtain
\bqa
&& F^{\tchi_i^0\tchi_j^0}_{-\mu,-\mu',-\tau,-\tau',}(\theta)=(-1)^{(\tau-\tau')}\eta_i\eta_j
  F^{\tchi_i^0\tchi_j^0}_{\mu \mu' \tau \tau'}(\theta)
  = (-1)^{(\tau-\tau')} F^{\tchi_j^0\tchi_i^0}_{-\mu', -\mu, -\tau', -\tau}(\theta) ~~,
  \label{chi00-CP} \\
&&  F^{\tchi_i^+\tchi_j^-}_{\mu\mu'\tau\tau'}(\theta)=
 F^{\tchi_j^+\tchi_i^-}_{-\mu', -\mu,- \tau',- \tau}(\theta) ~~.  \label{chipm-CP}
 \eqa

According to the  HCns theorem,  the helicity conserving  HC amplitudes,
which  satisfy
\bq
\mu+\mu'=\tau+\tau' ~~, \label{HC-constraint}
\eq
will be the only ones surviving asymptotically. These are
\bq
F^{ij}_{+-+-}(\theta)~~,~~ F^{ij}_{+--+}(\theta)~~,~~
F^{ij}_{-++-}(\theta)~~,~~ F^{ij}_{-+-+}(\theta)~,~\label{HC-amp}
\eq
constrained by (\ref{gg-Bose}) as
\bq
F^{ij}_{+--+}(\theta)= -F^{ij}_{-+-+}(\pi-\theta) ~~~,~~~
F^{ij}_{-++-}(\theta)=-F^{ij}_{+-+-}(\pi-\theta)~~. \label{HC-Bose-rel}
\eq
All helicity violating (HV) amplitudes, which by definition   violate (\ref{HC-constraint}),
 must vanish in the high energy limit.

On the basis of  (\ref{gg-Bose}-\ref{chipm-CP}),
 the independent    HC and HV  amplitudes for $gg\to \tchi_i^0\tchi_j^0$, may be chosen as
\bqa
HC & \Rightarrow &  F^{ij}_{-+-+}(\theta) ~~,~~ F^{ij}_{-++-}(\theta)~~, \nonumber \\
HV & \Rightarrow &  F^{ij}_{----}(\theta)~~,~~F^{ij}_{---+}(\theta)~~,~~
F^{ij}_{--+-}(\theta)~~,
~~F^{ij}_{--++}(\theta)~~, \nonumber \\
&& F^{ij}_{-+--}(\theta)~~,~~F^{ij}_{-+++}(\theta)~~. \label{Fchi00-ind}
\eqa
Correspondingly for $gg\to \tchi_i^+\tchi_j^-$, the independent  amplitudes
are chosen as
\bqa
HC & \Rightarrow &  F^{ij}_{-+-+}(\theta) ~~,~~ F^{ij}_{-++-}(\theta)~~,~~
F^{ij}_{+--+}(\theta) ~~,~~ F^{ij}_{+-+-}(\theta)~~,  \nonumber \\
HV & \Rightarrow &  F^{ij}_{----}(\theta)~,~F^{ij}_{---+}(\theta)~,~F^{ij}_{--+-}(\theta)~,
~F^{ij}_{--++}(\theta)~,~  F^{ij}_{-+--}(\theta)~,~F^{ij}_{-+++}(\theta),
\nonumber \\
&&   F^{ij}_{++--}(\theta)~,~F^{ij}_{++-+}(\theta)~,~F^{ij}_{+++-}(\theta)~,
~F^{ij}_{++++}(\theta)~,~  F^{ij}_{+---}(\theta)~,~F^{ij}_{+-++}(\theta).~~~
\label{Fchipm-ind}
\eqa\\

 The contributing   diagrams to the 1loop EW order, appearing in
Fig.\ref{diagrams-fig},   are calculated analytically, in
terms of Passarino-Veltman (PV) functions \cite{Veltman}.
To the contribution of  the triangular graphs
 $A,A',B,B'$  and the boxes $G,F,H $, we  also add the
 gluon  symmetrization (gSYM) contribution, obtained
 through the interchange
\bq
\mu \leftrightarrow \mu' ~~~,~~~ \cos\theta \leftrightarrow -\cos\theta  ~~~,~~~
\sin\theta \leftrightarrow -\sin\theta  ~~~.\label{SYM}
\eq
  The contributions of the graphs  $C,C',D$ are  already gluon-symmetrized.\\

Using these, the  Fortran code ggXXcode has been  constructed  giving
 $F_{\mu \mu'\tau\tau'}^{ij}(\sqrt{s},\theta)$,
in terms  of the c.m. energy in TeV and the c.m. angle
in radians \cite{code}. A factor $\alpha\alpha_s$ (together with $\delta^{ab}$)
is always removed from the amplitudes given by the code.
All input parameters are assumed at the electroweak scale, and
all mass-dimensional parameters are in TeV.
 The  quark masses of the first two generations are  neglected, as well as
CP violation,  so that all input parameters are real.
The PV functions  are calculated
 using the looptools subroutines  \cite{looptools1, looptools2}.
 The results of the codes are contained in output files  specified as ".dat".
 An accompanying Readme, fully explains the compilation  of the code \cite{code}.\\

\section{High energy behavior in MSSM}

Before addressing the   $gg\to \tchi^0_i\tchi^0_j,~\tchi^+_i\tchi^-_j$ asymptotic amplitudes;
we  recapitulate the corresponding  expressions for the $gg\to VV'$ amplitudes
$F^{VV'}_{\mu\mu'\tau\tau'}$, with  $VV'=ZZ$, $\gamma Z$, $\gamma\gamma$, $W^+W^-$,
and   $\tau,\tau'$
this time describing the $V,V'$ helicities \cite{ggVV}.

Abiding with the HCns theorem, only the  HC amplitudes can be non  vanishing asymptotically,
which implies that  $V,V'$ can be either both transverse, or both longitudinal.
For presenting   them,  it is convenient to define
\bq
\tilde \delta \left (\frac{x}{y} \right )
=-4\left [\ln^2 \left ({x  \over y}\right ) +\pi^2 \right ] ~~, \label{delta-gen}
\eq
 with $x$ and $y$ being complex, and the standard branches for the logarithms
 in the complex plane are used.
Using this, we define  \cite{ggVV}
\bqa
& & \delta_{+-+-}=\delta_{-+-+}\equiv \delta^t=
\tilde \delta \left(\frac{t+i\epsilon}{s+i\epsilon} \right )
\nonumber \\
&& =-4\left [\ln^2 \left ({2\over 1-\cos\theta}\right )
-i 2\pi \ln \left ({2\over 1-\cos\theta}\right ) \right ] ~~, \label{deltat} \\
& & \delta_{+--+}=\delta_{-++-} \equiv \delta^u =
\tilde \delta \left(\frac{u+i\epsilon}{s+i\epsilon} \right )
\nonumber \\
&& =-4\left [\ln^2 \left ({2\over 1+\cos\theta}\right )
-i 2\pi \ln \left ({2\over 1+\cos\theta}\right ) \right ] ~~, \label{deltau} \\
&& \delta_{++++}  =\delta_{----}\equiv \delta^{tu}=
 \tilde \delta \left(\frac{t+i\epsilon}{u +i\epsilon} \right )
=-4\left [\ln^2 \left ({1+\cos\theta \over 1-\cos\theta}\right )
+\pi^2 \right ] ~~, \label{deltapppp}
\eqa
which suffice to describe all asymptotic HC amplitudes  considered here. Note that
for asymptotic energies, the relations $t=-s(1-\cos\theta)/2$ and $u=-s(1+\cos\theta)/2$
are consistently used.

Using these,  the transverse gauge asymptotic  HC   amplitudes ($\tau\tau'\neq 0$),
become \cite{ggVV, gamgamZZ1, gamgamZZ2}
\bqa
&& F(gg\to ZZ)^{\rm as}_{\mu\mu' \tau\tau'} =\alpha\alpha_s
{(9-18s^2_W+20s^4_W)\over 24 s^2_Wc^2_W}
\delta_{\mu\mu' \tau\tau'} ~~, \nonumber \\
&& F(gg\to \gamma Z)^{\rm as}_{\mu\mu' \tau\tau'} = \alpha\alpha_s
{(9-20s^2_W)\over 24 s_Wc_W}
\delta_{\mu\mu' \tau\tau'} ~~, \nonumber \\
&& F(gg\to \gamma\gamma)^{\rm as}_{\mu\mu' \tau\tau'} = \alpha\alpha_s \, {5\over 6}
\, \delta_{\mu\mu' \tau\tau'} ~~, \nonumber \\
&& F(gg\to W^+W^-)^{\rm as}_{\mu\mu' \tau\tau'} = \alpha\alpha_s \, {3\over 8 s^2_W}
\, \delta_{\mu\mu' \tau\tau'}~~,  \label{HC-VVTTasym}
\eqa
while the  longitudinal ones ( $\tau=\tau'=0$) are\footnote{The quark masses
of the first two generations are neglected. }
\bqa
&& F(gg\to ZZ)^{\rm as}_{+-00}=F(gg\to ZZ)^{\rm as}_{-+00}=  \alpha\alpha_s
{(m^2_t+m^2_b)\over 16 s^2_W m^2_W} \Big \{{\delta^t(1-\cos\theta)\over 1+\cos\theta}+
{\delta^u(1+\cos\theta)\over1-\cos\theta}\Big \} ~~, \nonumber \\
&& F(gg\to W^+W^-)_{+-00}^{\rm as}= {\alpha\alpha_s\over 8 \swsq \mwsq }
\Big \{ {m_b^2\delta^t (1-\cos\theta)\over1+\cos\theta}+
  {m^2_t\delta^u (1+\cos\theta) \over 1-\cos\theta} \Big \}~~, \nonumber \\
 && F(gg\to W^+W^-)_{-+00}^{\rm as}= {\alpha\alpha_s\over 8 \swsq \mwsq }
\Big \{ {m_b^2\delta^u (1+\cos\theta)\over1-\cos\theta}+
  {m^2_t\delta^t (1-\cos\theta) \over 1+\cos\theta} \Big \}~~.  \label{HC-VVLLasym}
\eqa
For later use, we also deduce  from (\ref{HC-VVTTasym}) the asymptotic amplitudes
for the transverse $W_\mu$ and  $B_\mu$ gauge production
\bqa
&& F(gg\to W^+W^-)^{\rm as}_{\mu\mu' \tau\tau'} =
F(gg\to W^{(3)}W^{(3)})^{\rm as}_{\mu\mu' \tau\tau'}=
  {3\alpha\alpha_s \over 8 \swsq} \, \delta_{\mu\mu' \tau\tau'}~~, \nonumber \\
 && F(gg\to BB)^{\rm as}_{\mu\mu' \tau\tau'} =
 {11 \alpha\alpha_s \over 24 \cwsq} \, \delta_{\mu\mu' \tau\tau'}~~,
 \label{HC-gaugeasym}
\eqa
where (\ref{delta-gen}-\ref{deltapppp}) are used.

As seen from (\ref{HC-VVTTasym}, \ref{HC-VVLLasym}), the  HC asymptotic amplitudes
for $gg\to VV'$   are independent of the SUSY parameters.
For neutral gauge bosons, they  have  been first obtained
 in \cite{gamgamZZ1}, and the proof was extended to the $W^+W^-$ case in \cite{ggVV}.
 In both cases, the proof was   through  very
lengthy  computations.

For transverse gauge bosons, a much simpler way to verify
the  result (\ref{HC-gaugeasym}),
 is by studying  the supersymmetry-transformed  processes   describing the high energy
gluino   fusion to   gauginos; i.e.
$\tilde g \tilde g \to \tilde V \tilde V'$ with  $V V'=W^+W^-, ~W^{(3)}W^{(3)}, ~BB$.
At the 1loop EW order, the  asymptotically non vanishing HC amplitudes,
 are then found to  satisfy \cite{PVasym}
\bq
 (-1)^{\tilde \mu-\tilde \tau'}
F(\tilde g \tilde g \to \tilde V \tilde V')_{\tilde \mu \tilde \mu' \tilde\tau \tilde\tau'}
=F(gg \to VV')_{\mu\mu' \tau\tau'} ~~,
\label{gauge-gaugino-asym}
\eq
where $\tilde \mu, ~ \tilde \mu',~ \tilde\tau, ~ \tilde\tau' $
are the gluino and gaugino
helicities, which of course receive  half integers values.
As seen in (\ref{gauge-gaugino-asym}), most of the gauge and gaugino
asymptotic amplitudes,
are  identical. But for $\tilde \mu-\tilde \tau'=\pm 1$, sign differences appear which
must be related to the way we define   the gluino amplitudes,
where positive energy Dirac wave functions are used to  describe the first incoming
or outgoing fermionic particle, and negative energy Dirac wave functions
are used for describing
the second one.

The validity of (\ref{gauge-gaugino-asym}) for transverse $VV'$, is of course
a consequence of the fact that, at the 1loop EW level,
the high energy HC amplitudes for $gg\to VV'$
 are independent of  the SUSY breaking parameters, while in the corresponding
  $\tilde g \tilde g \to \tilde V \tilde V'$, the gaugino mixing is ignored.
  An  analogous result for $ug\to dW$ and its SUSY
 transformed process, could only be approximately true;
 since some dependence of the HC amplitudes
 on the SUSY breaking masses remains in this case, even
 at asymptotic energies \cite{ugsdWino, ugdW}.

In analogy to (\ref{gauge-gaugino-asym}),   the corresponding to
 asymptotic relations between the  longitudinal vector boson amplitudes
 and  the    gluino-to-higgsino ones, may be given by
  first  noting  that
\bqa
&& F( \tilde g\tilde g \to \tilde H_1^0 \tilde H_1^0)_{\pm\mp\pm\mp}=
\frac{\alpha \alpha_s}{8}\,
\frac{m_b^2}{\mwsq \swsq \cos^2\beta}\, \frac{(1-\cos\theta)}{(1+\cos\theta)}\, \delta^t ~~,
\nonumber \\
&& F( \tilde g\tilde g \to \tilde H_1^0 \tilde H_1^0)_{\mp\pm\pm\mp}=
-\frac{\alpha \alpha_s}{8}\,
\frac{m_b^2}{\mwsq \swsq \cos^2\beta}\, \frac{(1+\cos\theta)}{(1-\cos\theta)}\, \delta^u ~~,
\nonumber \\
&& F( \tilde g\tilde g \to \tilde H_2^0 \tilde H_2^0)_{\pm\mp\pm\mp}=
\frac{\alpha \alpha_s}{8}\,
\frac{m_t^2}{\mwsq \swsq \sin^2\beta}\, \frac{(1-\cos\theta)}{(1+\cos\theta)}\, \delta^t ~~,
\nonumber \\
&& F( \tilde g\tilde g \to \tilde H_2^0 \tilde H_2^0)_{\mp\pm\pm\mp}=
-\frac{\alpha \alpha_s}{8}\,
\frac{m_t^2}{\mwsq \swsq \sin^2\beta}\, \frac{(1+\cos\theta)}{(1-\cos\theta)}\, \delta^u ~~,
\nonumber \\
&& F( \tilde g\tilde g \to \tilde H_1^+ \tilde H_1^-)_{+-+-}=
\frac{\alpha \alpha_s}{8}\,
\frac{m_b^2}{\mwsq \swsq \cos^2\beta}\, \frac{(1-\cos\theta)}{(1+\cos\theta)}\, \delta^t ~~,
\nonumber \\
&& F( \tilde g\tilde g \to \tilde H_2^+ \tilde H_2^-)_{+-+-}=
\frac{\alpha \alpha_s}{8}\,
\frac{m_t^2}{\mwsq \swsq \sin^2\beta}\, \frac{(1-\cos\theta)}{(1+\cos\theta)}\, \delta^t ~~,
\label{gluino-higgsino-asym}
\eqa
at 1loop EW order, where (\ref{deltat},\ref{deltau}) are used.
Comparing these to  (\ref{HC-VVLLasym})
we obtain the asymptotic relations
\bqa
& F(gg\to ZZ)^{\rm as}_{+-00}&=\frac{\sin^2\beta}{2}
\Big [F( \tilde g\tilde g \to \tilde H_2^0 \tilde H_2^0)_{+-+-}-
F( \tilde g\tilde g \to \tilde H_2^0 \tilde H_2^0)_{+--+} \Big ] \nonumber \\
&& + \frac{\cos^2\beta}{2}
\Big [F( \tilde g\tilde g \to \tilde H_1^0 \tilde H_1^0)_{+-+-}-
F( \tilde g\tilde g \to \tilde H_1^0 \tilde H_1^0)_{+--+} \Big ]~, \nonumber \\
& F(gg\to ZZ)^{\rm as}_{-+00}&=\frac{\sin^2\beta}{2}
\Big [F( \tilde g\tilde g \to \tilde H_2^0 \tilde H_2^0)_{-+-+}-
F( \tilde g\tilde g \to \tilde H_2^0 \tilde H_2^0)_{-++-} \Big ] \nonumber \\
&& + \frac{\cos^2\beta}{2}
\Big [F( \tilde g\tilde g \to \tilde H_1^0 \tilde H_1^0)_{-+-+}-
F( \tilde g\tilde g \to \tilde H_1^0 \tilde H_1^0)_{-++-} \Big ]~, \nonumber \\
& F(gg\to W^+W^-)^{\rm as}_{+-00}&=\cos^2\beta
F( \tilde g\tilde g \to \tilde H_1^+ \tilde H_1^-)_{+-+-}
-\sin^2\beta F( \tilde g\tilde g \to \tilde H_2^+ \tilde H_2^-)_{+--+} ~, \nonumber \\
& F(gg\to W^+W^-)^{\rm as}_{-+00}&=-\cos^2\beta
F( \tilde g\tilde g \to \tilde H_1^+ \tilde H_1^-)_{-++-} \nonumber \\
&& +\sin^2\beta F( \tilde g\tilde g \to \tilde H_2^+ \tilde H_2^-)_{-+-+} ~,
\label{Goldstone-higgsino-asym}
\eqa
in agreement with the SUSY transformations relating the higgsinos to
the $G^{\pm},~G^0$ Goldstone bosons, and   the longitudinal
EW vector bosons, through the equivalence theorem \cite{couplings}.
Note that in contrast to   (\ref{gauge-gaugino-asym}, \ref{HC-gaugeasym})
which are  independent of any symmetry  breaking
parameter, some  symmetry breaking appears
in  (\ref{Goldstone-higgsino-asym}), contained in the angle $\beta$.

  The actual form of  (\ref{gauge-gaugino-asym}) and (\ref{Goldstone-higgsino-asym})
  is an impressive example of how contrived  supersymmetry  can be.
  Without SUSY, we would naively expect that the angular dependencies
 of  amplitudes involving particle of different spin, cannot be the same;
 particularly for the $F_{+-+-}$
  amplitudes in   (\ref{gauge-gaugino-asym}),
 where  the\footnote{Remember that all helicities in the l.h.s. of
 (\ref{gauge-gaugino-asym}) are $\pm 1/2$, while in the r.h.s are $\pm 1$. }
  $d^J$ functions involved in the partial
 wave expansion  are  very different \cite{JW}. It is amusing to see how SUSY manages
 to keep the amplitudes unchanged, after  all partial waves are summed over.
 A genuine SUSY property indeed, not shared by any other symmetry in particle physics!  \\

We next turn to the asymptotic amplitudes for
$gg\to \tchi^0_i\tchi^0_j,~\tchi^+_i\tchi^-_j$; with  $\tau,\tau'$ now describing  the
neutralino and chargino helicities. These final states are the supersymmetric
transformed of the final states in (\ref{HC-VVTTasym}, \ref{HC-VVLLasym}).
As expected, only the HC amplitudes  survive asymptotically, while the  HV ones
vanish at very high energies.

The contributions of the diagrams of Fig.\ref{diagrams-fig},
to the various  HV amplitudes at high energies \cite{PVasym}, are as follows:
\begin{itemize}

\item
The HV amplitudes for  $\mu=\mu'$  and $\tau=\tau'$, vanish  asymptotically
like $\sim m/\sqrt{s}$, without any  logarithmic corrections.

\item
The HV amplitudes for    $\mu=\mu'$ and  $\tau=-\tau'$ are  strongly  suppressed
like $\sim m^2/s$, again with no   logarithmic corrections. These amplitudes receive
their main contributions from
the boxes $F,G,H$.

\item
 Finally, the HV amplitudes for  $\mu=-\mu'$  and $\tau=\tau'$,
 receive high energy contributions
of the form
\bq
 -{\alpha\alpha_s\over4}\left ({2m_q\over\sqrt{s}}\right )
\left [\ln^2\left ({t\over m^2_q} \right)
+\ln^2\left({u\over m^2_q} \right )-\ln^2\left({s \over m^2_q} \right) \right] ~~,
\label{pmtautau}
\eq
arising  from  third generation quark-squark boxes, where $m_q$ is the top
or bottom mass.
These are the slowest vanishing HV amplitudes, at high energies.\\

\end{itemize}

We next turn to the HC amplitudes  for   $gg\to \tchi_i\tchi_j$,
which necessarily  satisfy
 $\mu=-\mu'$ and $\tau=-\tau'$,  and behave  asymptotically
like  angular dependent "constants", mainly feeded by
the F, G and H boxes of Fig.\ref{diagrams-fig}. In all cases,  the approach
 to the asymptotic  values   is determined by power laws, possibly corrected
 by logarithms. For  charginos these are
\bqa
&&F(gg\to \tchi^+_i\tchi^-_j)_{-+-+}^{as}=
\frac{\alpha\alpha_s}{8\swsq} {\delta^t(1-\cos\theta)\over\sin\theta}
\left \{3 Z^{+*}_{1i}Z^{+}_{1j}
+Z^{+*}_{2i}Z^{+}_{2j}{m^2_t\over \mwsq\sin^2\beta} \right \} ~, \nonumber \\
&&F(gg\to \tchi^+_i\tchi^-_j)_{+-+-}^{as}= -
\frac{\alpha\alpha_s}{8\swsq} {\delta^t(1-\cos\theta)\over\sin\theta}
\left \{3 Z^{-}_{1i}Z^{-*}_{1j}
+Z^{-}_{2i}Z^{-*}_{2j}{m^2_b\over \mwsq\cos^2\beta} \right \} ~, \nonumber \\
&&F(gg\to \tchi^+_i\tchi^-_j)_{+--+}^{as}= -
\frac{\alpha\alpha_s}{8\swsq} {\delta^u(1+\cos\theta)\over\sin\theta}
\left \{3 Z^{+*}_{1i}Z^{+}_{1j}
+Z^{+*}_{2i}Z^{+}_{2j}{m^2_t\over \mwsq\sin^2\beta} \right \} ~, \nonumber \\
&&F(gg\to \tchi^+_i\tchi^-_j)_{-++-}^{as}=
\frac{\alpha\alpha_s}{8\swsq} {\delta^u(1+\cos\theta)\over\sin\theta}
\left \{3 Z^{-}_{1i}Z^{-*}_{1j}
+Z^{-}_{2i}Z^{-*}_{2j}{m^2_b\over \mwsq\cos^2\beta} \right \} ~, \label{HC-chipmasym}
\eqa
while for neutralinos
\bqa
&&F(gg \to \tchi^0_j\tchi^0_i)_{-+-+}^{as}=F(gg \to \tchi^0_i\tchi^0_j)_{-+-+}^{as  *}
=-F(gg\to \tchi^0_i\tchi^0_j)_{+-+-}^{as}  =
\alpha\alpha_s {\delta^t(1-\cos\theta)\over\sin\theta}
\nonumber \\
&& \cdot \Big \{  Z^{N}_{1i}Z^{N*}_{1j}{11\over24c^2_W}
 +Z^N_{2i}Z^{N*}_{2j}{3\over8s^2_W}
+Z^{N}_{3i}Z^{N*}_{3j}{m^2_b\over8s^2_W m^2_W\cos^2\beta}
+Z^{N}_{4i}Z^{N*}_{4j}{m^2_t\over8s^2_W m^2_W\sin^2\beta}\Big \} ~~, \nonumber \\
&&F(gg \to \tchi^0_i\tchi^0_j)_{-++-}^{as}=-F(gg\to \tchi^0_i\tchi^0_j)_{+--+}^{as *}
=-F(gg\to \tchi^0_j\tchi^0_i)_{+--+}^{as}
=\alpha\alpha_s {\delta^u(1+\cos\theta)\over\sin\theta}
\nonumber \\
&& \cdot \Big \{  Z^{N}_{1i}Z^{N*}_{1j}{11\over24c^2_W}
 +Z^N_{2i}Z^{N*}_{2j}{3\over8s^2_W}
+Z^{N}_{3i}Z^{N*}_{3j}{m^2_b\over8s^2_W m^2_W\cos^2\beta}
+Z^{N}_{4i}Z^{N*}_{4j}{m^2_t\over8s^2_W m^2_W\sin^2\beta}\Big \} ~~, \label{HC-chi00asym}
\eqa
where $Z^-_{\alpha i}, Z^+_{\alpha i} $ and $Z^N_{\alpha i}$ are the chargino and neutralino
mixing matrices respectively \cite{couplings}.

Technically, the  asymptotic expressions (\ref{HC-chipmasym},\ref{HC-chi00asym})
  were first obtained by studying   gluon-fusion
  to pairs of bino, wino and higgsino components.
  Afterwards, the final  expressions for any physical $gg\to\tchi_i\tchi_j$ process
  were  constructed,   by introducing  the corresponding $Z$ mixing-matrix elements.

The asymptotic expressions for the dimensionless differential cross sections $\tilde \sigma$
defined in (\ref{sigmatilde}), for chargino-neutralino
and vector-boson production are given by
\bqa
 \tilde \sigma(gg\to \tchi_i\tchi_j)^{as}  & \equiv &
 \frac{\sum_{HC}|F(gg\to \tchi_i\tchi_j)^{as}_{\mu\mu'\tau\tau'}|^2}
{\alpha^2 \alpha_s^2}~~~, \nonumber \\
 \tilde \sigma(gg\to VV')^{as}  & \equiv &
 \frac{\sum_{HC}|F(gg\to VV')^{as}_{\mu\mu'\tau\tau'}|^2}
{\alpha^2 \alpha_s^2}~~~,  \label{sigmatilde-as}
\eqa
where the summation is over the asymptotic HC amplitudes in
(\ref{HC-VVTTasym}, \ref{HC-VVLLasym}, \ref{HC-chipmasym}, \ref{HC-chi00asym}).

 As seen from (\ref{HC-chipmasym},\ref{HC-chi00asym}),
 the sizes of the gaugino  and  higgsino parts of the HC amplitudes
 are solely determined by the
 $Z$ mixing-matrix elements, multiplied by coefficients which, for large $\tan\beta$,
 acquire similar magnitudes; compare the terms   within
  the curly brackets in (\ref{HC-chipmasym},\ref{HC-chi00asym}).
  So the SUSY benchmark dependence  of the asymptotic HC amplitudes
  and    $\tilde \sigma(gg\to \tchi_i\tchi_j)^{as}$   will  be
controlled by the sizes of the Z-matrix elements and the $\beta$-angle.

In the "inclusive" asymptotic cross section
$\sum_{ij}\tilde\sigma(gg\to \tchi_i \tchi_j)^{as}$ though,
where all chargino or neutralino final states  are summed over,
the unitarity of the Z-matrices eliminates the Z-dependence, so that
the  angle $\beta$ is the sole SUSY parameter on which this quantity  depends. \\

Combining (\ref{sigmatilde}, \ref{sigmatilde-as}),  this
 implies that the energy- and angle-dependent quantities
\bqa
&&  \frac{\tilde\sigma(gg\to W^+W^-)}{\tilde\sigma(gg\to W^+W^-)^{as}}
 ~,~ \frac{\tilde\sigma(gg\to ZZ)}{\tilde\sigma(gg\to ZZ)^{as}}
~,~  \frac{\tilde\sigma(gg\to \gamma Z)}{\tilde\sigma(gg\to \gamma Z)^{as}}
~,~  \frac{\tilde\sigma(gg\to \gamma \gamma )}{\tilde\sigma(gg\to \gamma \gamma )^{as}} ~~,
  \label{Rsigma1} \\
&& \frac{\sum_{ij}\tilde\sigma(gg\to \tchi^+_i\tchi^-_j)}
{\sum_{ij}\tilde\sigma(gg\to \tchi^+_i\tchi^-_j)^{as}} ~~,~~
  \frac{\sum_{ij}\tilde\sigma(gg\to \tchi_i^0\tchi_j^0)}
{\sum_{ij}\tilde\sigma(gg\to \tchi^0_i \tchi^0_j)^{as}} ~~, \label{Rsigma2} \\
&& \frac{\tilde\sigma(gg\to \tchi^+_i\tchi^-_j)}{\tilde\sigma(gg\to \tchi^+_i\tchi^-_j)^{as}}
~~,~~ \frac{\tilde\sigma(gg\to \tchi_i^0\tchi_j^0)}
{\tilde\sigma(gg\to \tchi^0_i \tchi^0_j)^{as}}
~~, \label{Rsigma3}
\eqa
should all approach 1, as the energy reaches  sufficient process dependent  values.

 Note  that the  denominators $\tilde \sigma^{as}$ in (\ref{Rsigma1}) for gauge production,
 only depend on the gauge couplings  and are  independent
 of the SUSY benchmark; compare (\ref{HC-VVTTasym}, \ref{HC-VVLLasym}).
 As already said,  the only SUSY dependence  of the denominators in  (\ref{Rsigma2}),
 is  contained in the angle $\beta$; while the SUSY dependence of
 the denominators of the  two quantities in    (\ref{Rsigma3}),
 also involves  Z-matrix elements.

 In the next Section we will  see how the various quantities in
 (\ref{Rsigma1}, \ref{Rsigma2}, \ref{Rsigma3})   approach unity,  as  the energy increases. \\

\section{Numerical results  }

\begin{table}[hbt]
\begin{center}
{ Table 1: Asymptotic  HC amplitudes   for $gg \to \tchi_1\tchi_2$
 divided by $\alpha \alpha_s$,\\
and  $\tilde \sigma(gg \to \tchi_1\tchi_2)^{as}$ at $\theta=60^o$,
in $SPS1a'$ \cite{SPA1}.  }\\
  \vspace*{0.3cm}
\begin{small}
\begin{tabular}{||c|c||c|c||}
\hline \hline
\multicolumn{2}{||c||}{$gg \to \tchi^0_1 \tchi^0_2 $} &
\multicolumn{2}{c||}{$gg \to \to \tchi^+_1 \tchi^-_2 $} \\
 \hline
 $F_{-+-+}(\theta)$ & $ 0.21-i 0.94 $  &   $F_{-+-+}(\theta)$ & $ 0.8-i 3.6 $   \\
  $F_{+-+-}(\theta)$ & $ -0.21 +i 0.94 $ &   $F_{+-+-}(\theta)$ & $ 2.1-i 9.6 $     \\
   $F_{+--+}(\theta)$ & $ - 0.027+ i 0.59 $  &   $F_{+--+}(\theta)$ & $-0.10 +i 2.3 $   \\
    $F_{-++-}(\theta)$ & $ 0.027 - i 0.59 $ &   $F_{-++-}(\theta)$ & $-0.27 + i 6.0 $    \\
 \hline  \hline
\end{tabular}
 \end{small}
\end{center}
\end{table}

In this section we give numerical illustrations for the $gg\to \tchi_1\tchi_2$ amplitudes.
As already said,  all   HV amplitudes
vanish asymptotically, while the asymptotic HC amplitudes are given
by (\ref{HC-chipmasym}, \ref{HC-chi00asym}). Using these, we give  in Table 1,
the   HC asymptotic amplitudes at $\theta=60^o$ for the benchmark $SPS1a'$ \cite{SPA1}.
Needles to say, that these results fully agree with those obtained from
the complete 1loop code at high energy  \cite{code}.\\

We next turn to asymptotic  dimensionless differential
cross sections $\tilde \sigma^{as}$ defined in (\ref{sigmatilde-as}) and
compare  the asymptotic production of the  specific neutralino
or chargino pair with  $(i=1, ~j=2)$),
with the case where summation over all charginos or neutralinos is done.
As said above, in the first case strong SUSY benchmark dependence appears, while
in the second case, the only SUSY dependence is through the angle $\beta$.
As SUSY benchmarks we select five constrained MSSM models defined in Table 2,
where the grand  scale  $\tan\beta$ varies from 10 to 50; while  the bino,  wino
and  higgsino components of the lightest  neutralinos and chargino cover
a wide range of possibilities.

\begin{table}[h]
\begin{center}
{ Table 2: Input  parameters at the grand scale, for five constrained\\
 MSSM benchmark models with   $\mu>0$. All dimensional parameters  in GeV. }\\
  \vspace*{0.3cm}
\begin{small}
\begin{tabular}{||c|c|c|c|c|c||}
\hline \hline
  & $SPS1a'$ \cite{SPA1} &   mSP4 \cite{Baer2}& BBSSW \cite{Baer1}
  & AD1 \cite{Arnowitt}  & BKPU \cite{Baer4}  \\ \hline
 $m_{1/2}$ & 250   & 137&900 & 900  & 2900   \\
 $m_0$ & 70   & 1674& 4716 & 400 & 8700   \\
 $A_0$ & -300&  1985  & 0 & 0 & 0   \\
$\tan\beta$ & 10  & 18.6& 30 & 40  & 50  \\
  \hline \hline
\end{tabular}
 \end{small}
\end{center}
\end{table}

In addition, we  consider  the  non-universal AD2 model suggested in  \cite{Arnowitt},
 again characterized by $\mu>0$, but having  unequal  Higgs masses
 at the grand scale. Its  defining  high scale parameters are
\bqa
 {\rm AD2-model }&& M_1=M_2=M_3=A_0=420  ~,~  \tan\beta=40 ~,~
\nonumber \\
&& m_0=500 ~,~ m_{H_u}^2=6\cdot 10^{5}~,~ m_{H_d}^2=3.6\cdot 10^{5}~~, \label{AD2}
\eqa
where all dimensional parameters are in GeV.
Note that the grand  scale values for $\tan\beta$  in  AD1 and AD2, are the same.

The  lightest neutralino $\tchi^0_1$ in these models is mostly
a bino   for $SPS1a'$, mSP4, BBSSW, AD1;  and a higgsino for AD2,   BKPU.
Correspondingly $\tchi^0_2$ is mostly a wino for $SPS1a'$, mSP4,  AD1;
and a higgsino  for BBSSW, AD2, BKPU. Finally $\tchi^+_1$ is mainly a wino for
$SPS1a'$, mSP4, AD1; and a higgsino for BBSSW, AD2, BKPU.

The corresponding results for $\tilde \sigma^{as}$ are given in Table 3,
where of course the EW scale of the various SUSY parameters are used.

\begin{table}[h]
\begin{center}
{ Table 3: Asymptotic dimensionless cross sections for
$\tilde \sigma(gg\to \tchi_1\tchi_2)^{as}$ and
$\sum_{ij} \tilde\sigma(gg\to \tchi_i\tchi_j)^{as}$ at $\theta=60^o$, for the benchmarks
of Table 2 and AD2. }\\
  \vspace*{0.3cm}
  \hspace*{-0.3cm}
\begin{small}
\begin{tabular}{||c|c|c|c|c||}
\hline \hline
  & $\tilde \sigma(gg\to \tchi_1^0\tchi_2^0)^{as}$
  & $\sum_{ij} \tilde\sigma(gg\to \tchi_i^0\tchi_j^0)^{as}$
  & $\tilde \sigma(gg\to \tchi_1^+\tchi_2^-)^{as}$
  & $\sum_{ij} \tilde\sigma(gg\to \tchi_i^+\tchi_j^-)^{as}$   \\
  \hline
  $SPS1a'$ & 2.2   & 11294  & 156 & 6961   \\
 mSP4 & 1.2    & 11506 & 31  & 7066  \\
  BBSSW  & 66 & 13488   & 62   & 8058 \\
AD1   &  0.7 & 19809  & 38  & 11218  \\
AD2   & 35  & 19791  & 77&  11209   \\
BKPU   & 776  & 31617   & 4.6  & 17122  \\
  \hline \hline
\end{tabular}
 \end{small}
\end{center}
\end{table}

As seen from Table 3, the results for $\tilde\sigma(gg\to \tchi_1\tchi_2)^{as}$
vary from model to model, in a more or less random  way, depending on the many SUSY
parameters affecting the relevant Z-matrix elements.
In contrast to this,  the benchmark dependence largely  disappears in
 $ \sum_{ij} \tilde\sigma(gg\to \tchi_i\tchi_j)^{as}$,
 with only the dependence on the EW $\beta$-value
 remaining\footnote{Note that the slight differences between the AD1 and AD2
 results in the 3rd and 5th column of Table 3, solely come from
  small  differences between the EW scale $\tan\beta$ values. At the grand scale,
 $\tan\beta=40$, for both these models. };
 compare (\ref{HC-chipmasym}, \ref{HC-chi00asym}) and (\ref{sigmatilde-as})
 In all cases, irrespective of the nature of the lightest neutralinos or charginos,
 the  dimensionless cross sections satisfy
 \bqa
\sum_{ij}\tilde \sigma(gg \to \tchi^0_i \tchi^0_j)
 & \gg & \tilde \sigma(gg \to \tchi^0_1 \tchi^0_2) ~~,  \nonumber \\
\sum_{ij}\tilde \sigma(gg \to \tchi^+_i \tchi^-_j)
 & \gg & \tilde \sigma(gg \to \tchi^+_1 \tchi^-_2) ~~, \label{sigmatilde-ineq}
\eqa
at asymptotic energies  \cite{SPA1,SPA2, Baer1,Baer2, Baer3, Baer4, Arnowitt}.
It may be worthwhile to remark though, that for the higgsino $\tchi^0_1, \tchi^0_2 $ cases
of the models AD2 and BKPU, these inequalities are somewhat less strong.
 The physical reason for these strong inequalities should be due to the orthogonality of the $(\tchi_1, \tchi_2)$ states and  the $SU(2)$ EW
 gauge symmetry of MSSM.  This may  be realized  by
contemplating on the structure of (\ref{HC-chipmasym},\ref{HC-chi00asym}),
which also leads to the conclusion that at least some of the  diagonal production
cross sections $\tilde \sigma(gg \to \tchi_i \tchi_i)$, should be
comparable to the summed quantities in the l.h.s. of (\ref{sigmatilde-ineq}).
Exactly which, is of course  model-dependent. For $SPS1a'$ we have  checked
these statements numerically. \\

\begin{figure}[p]
\vspace*{-1cm}
\[
\epsfig{file=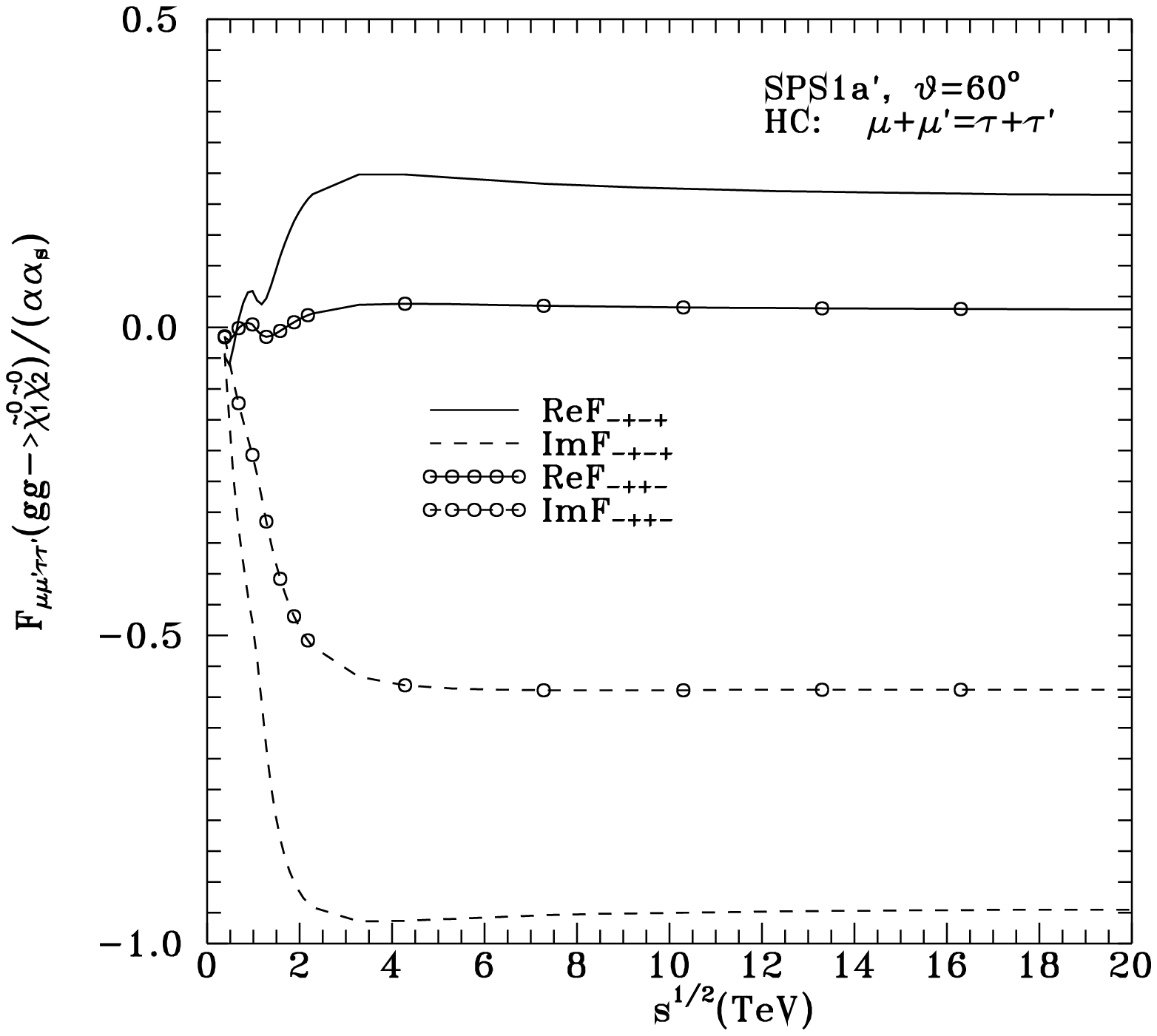, height=6.5cm}
\vspace*{0.5cm}
\]
\[
\epsfig{file=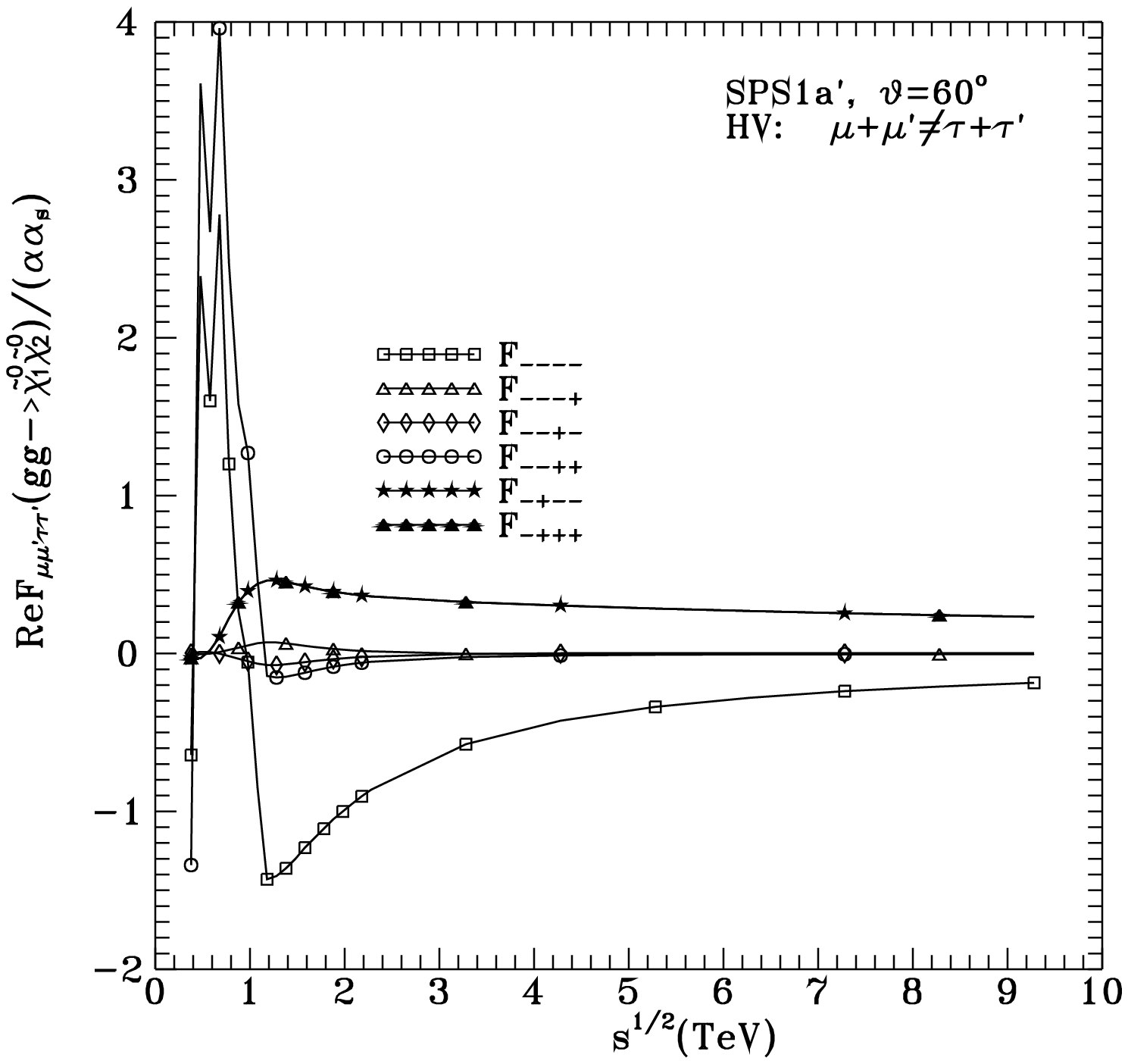, height=6.5cm}\hspace{1.cm}
\epsfig{file=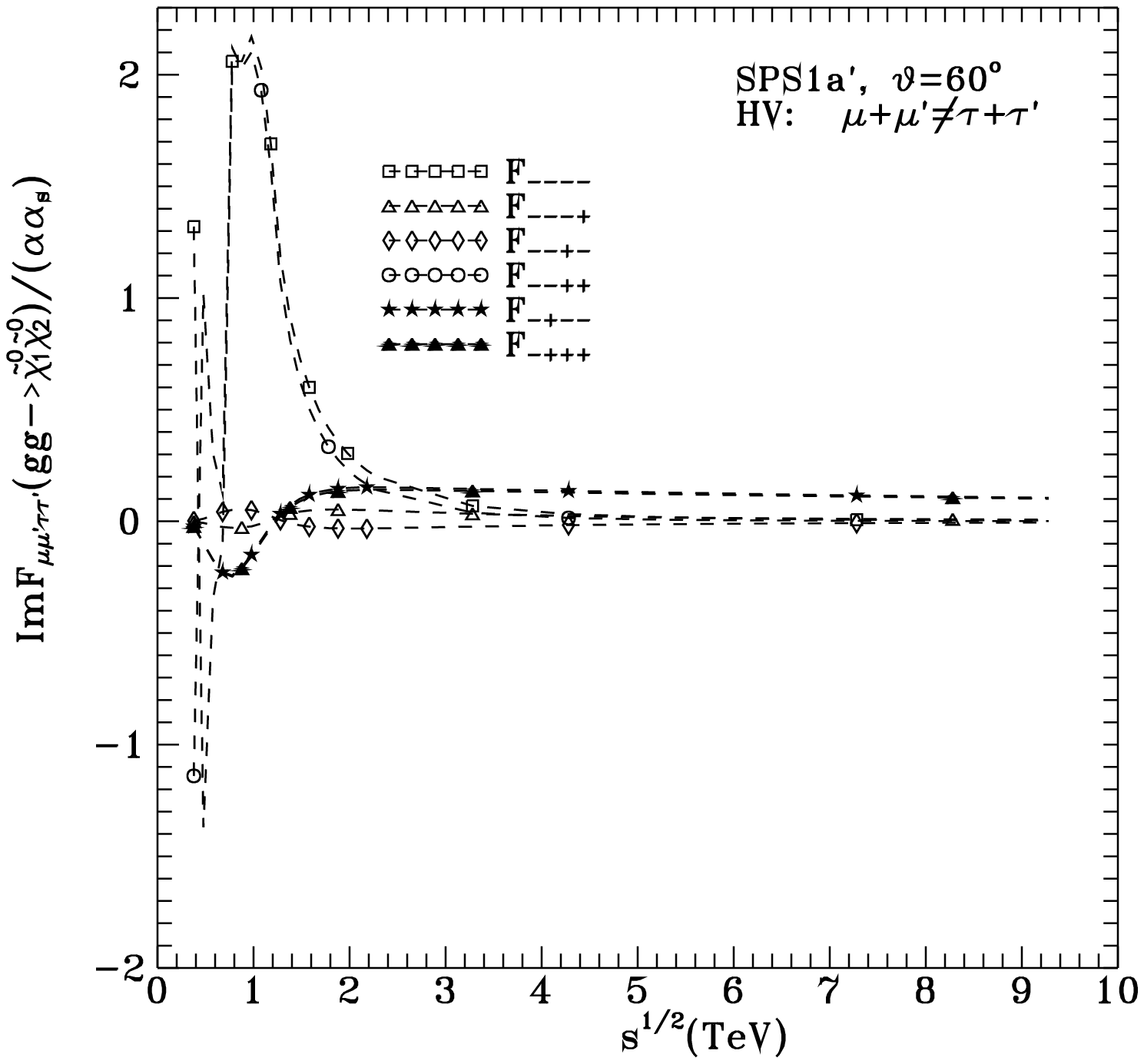,height=6.5cm}
\]
\caption[1]{Independent helicity amplitudes  for   $gg \to \tchi^0_1 \tchi^0_2 $
defined in (\ref{Fchi00-ind}), at $\theta=60^o$ in $SPS1a'$ \cite{SPA1}.
The upper panel  gives the real and imaginary parts of HC amplitudes, while the lower panels
indicate  the real and imaginary parts of the HV amplitudes. }
\label{amp-chi0012-SPA-fig}
\end{figure}

\begin{figure}[p]
\vspace*{-1cm}
\[
\epsfig{file=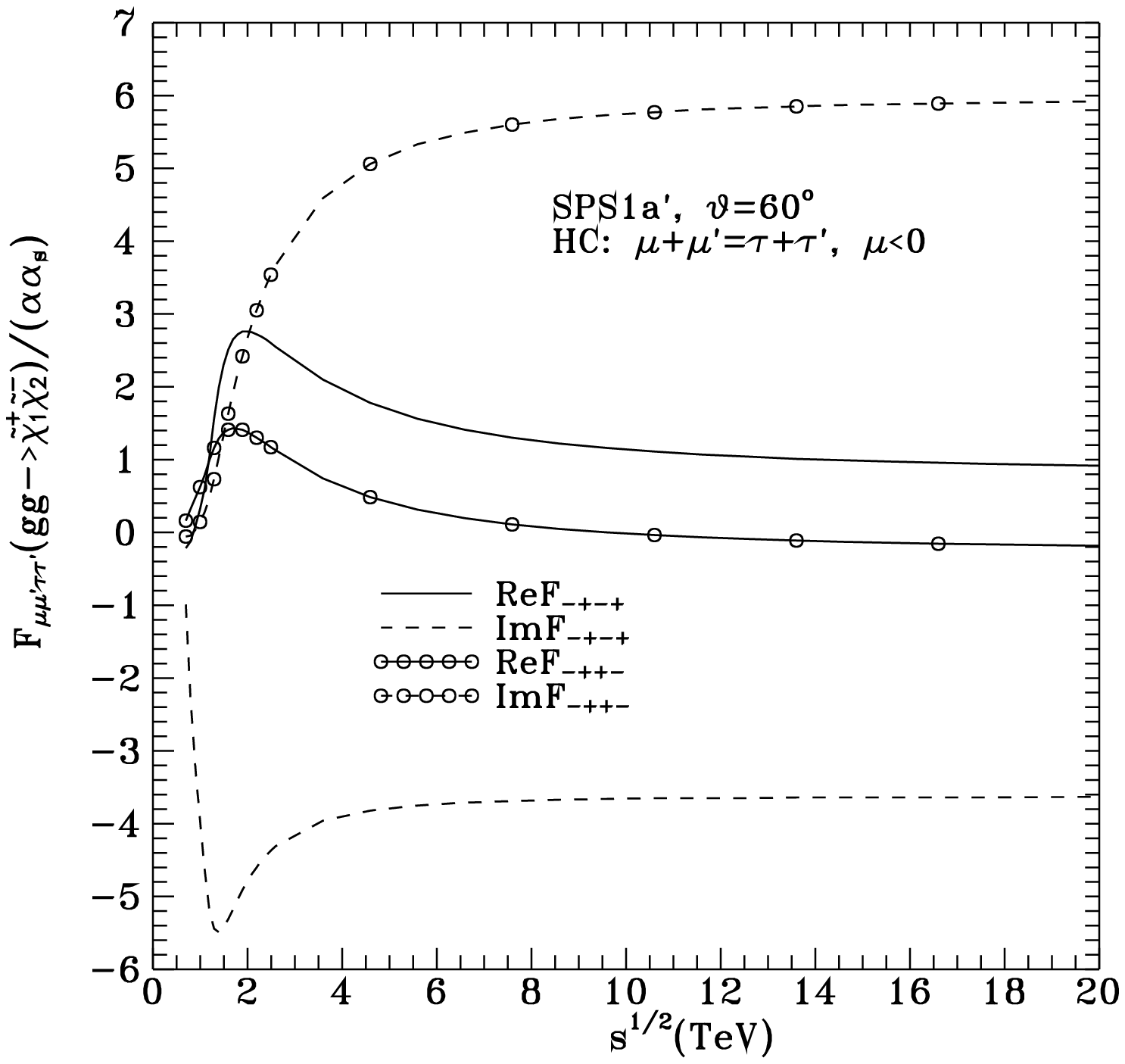, height=6.cm}\hspace{1.cm}
\epsfig{file=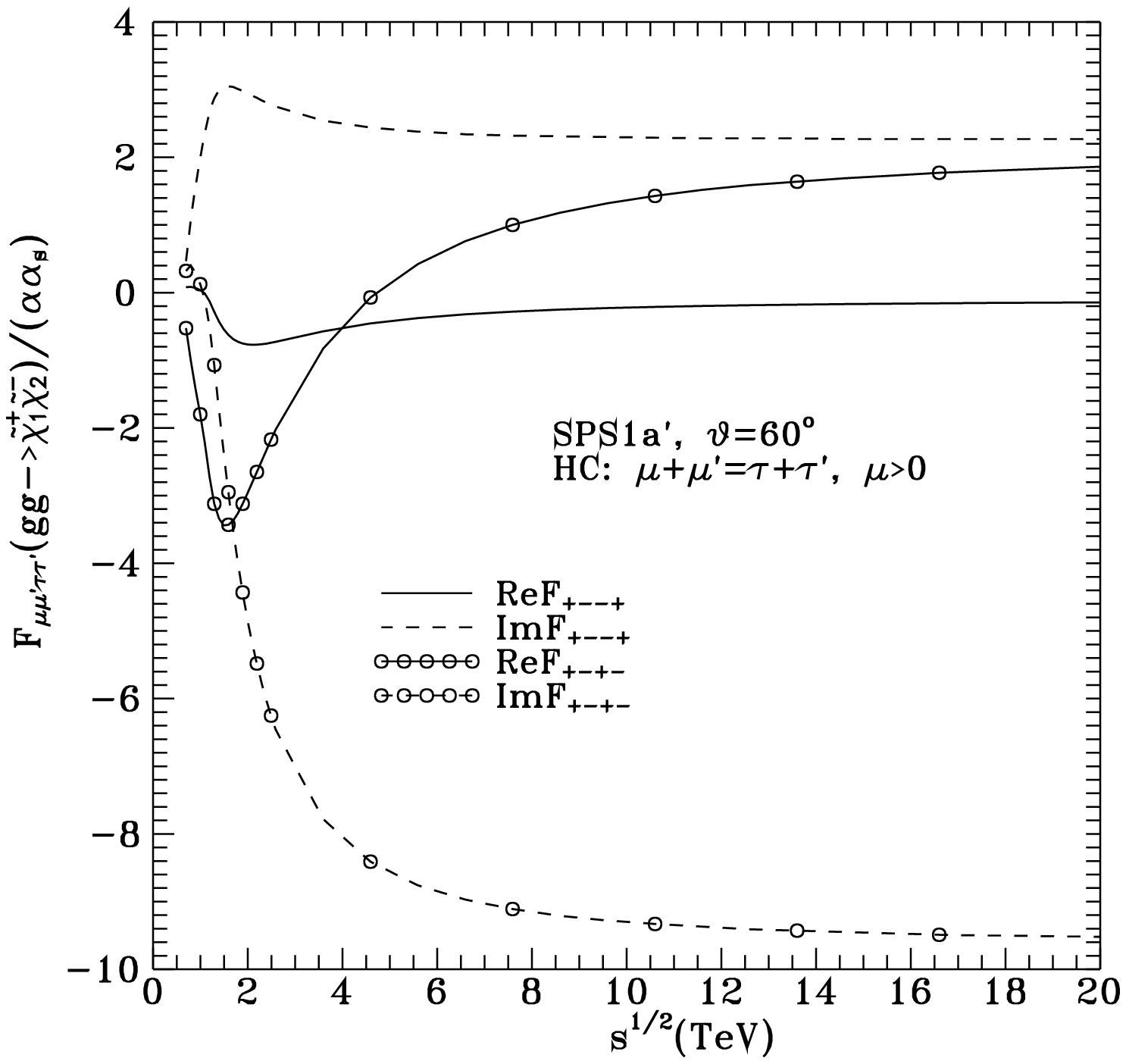,height=6.cm}
\]
\[
\epsfig{file=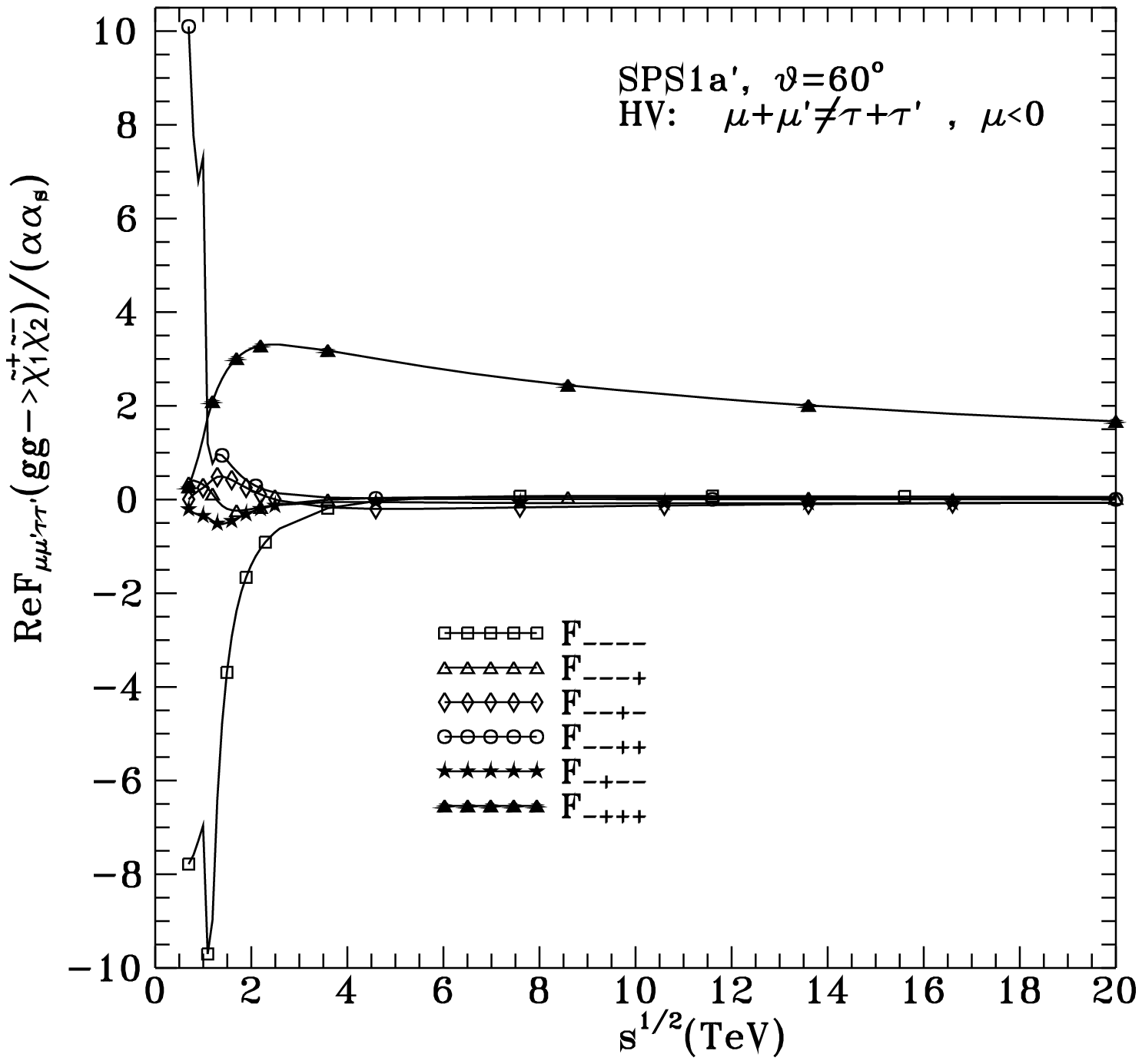, height=6.cm}\hspace{1.cm}
\epsfig{file=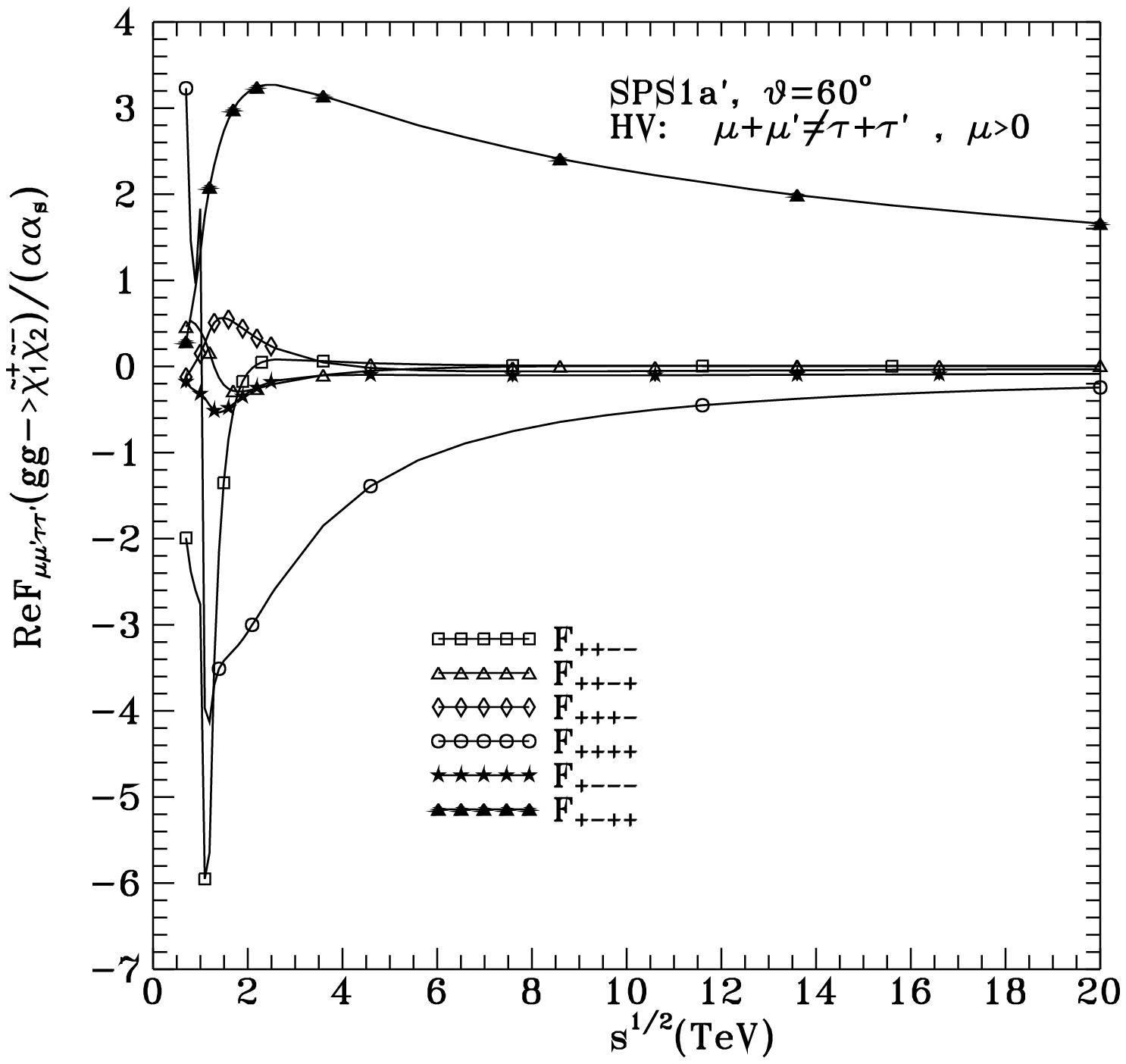,height=6.cm}
\]
\[
\epsfig{file=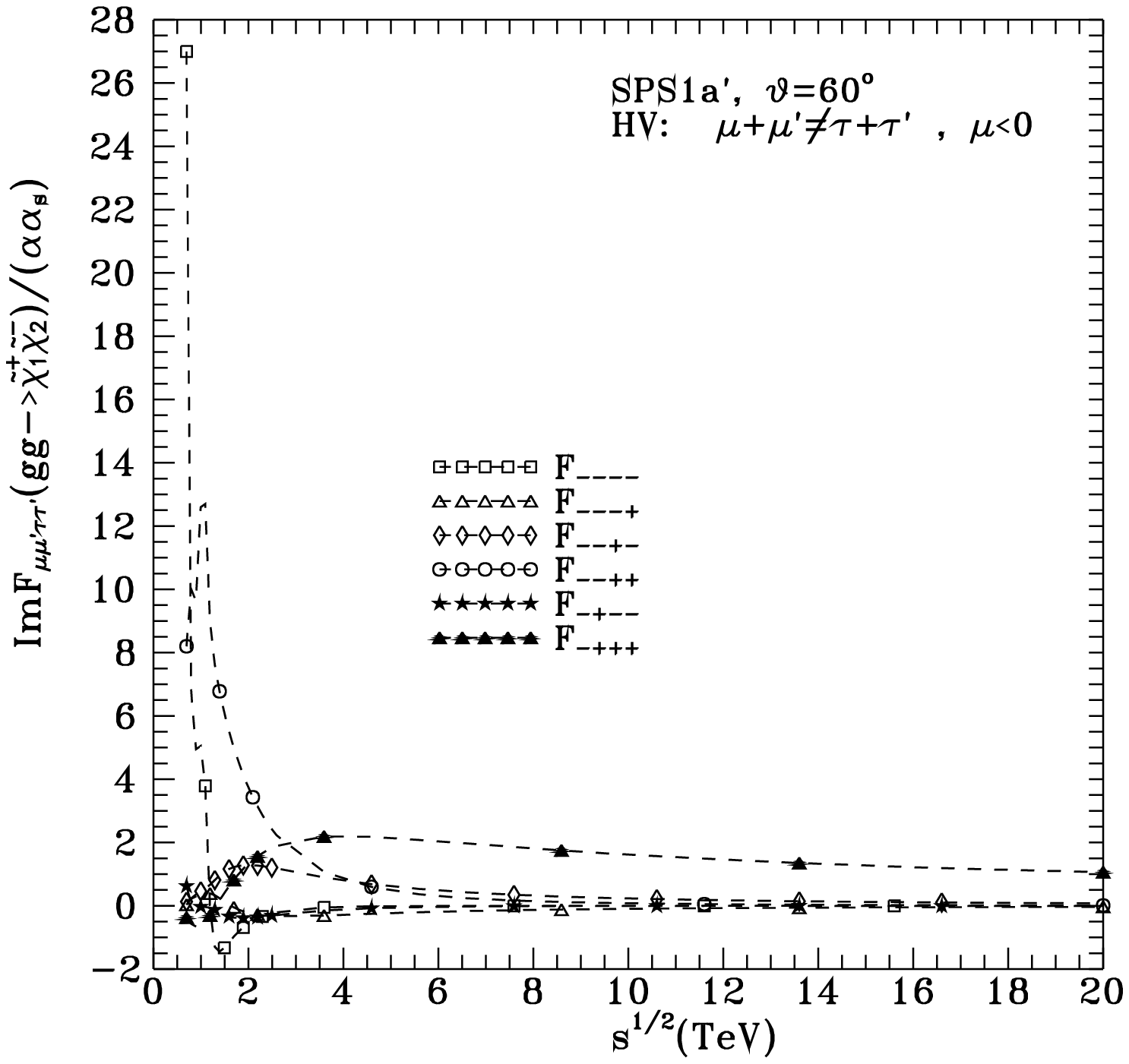, height=6.cm}\hspace{1.cm}
\epsfig{file=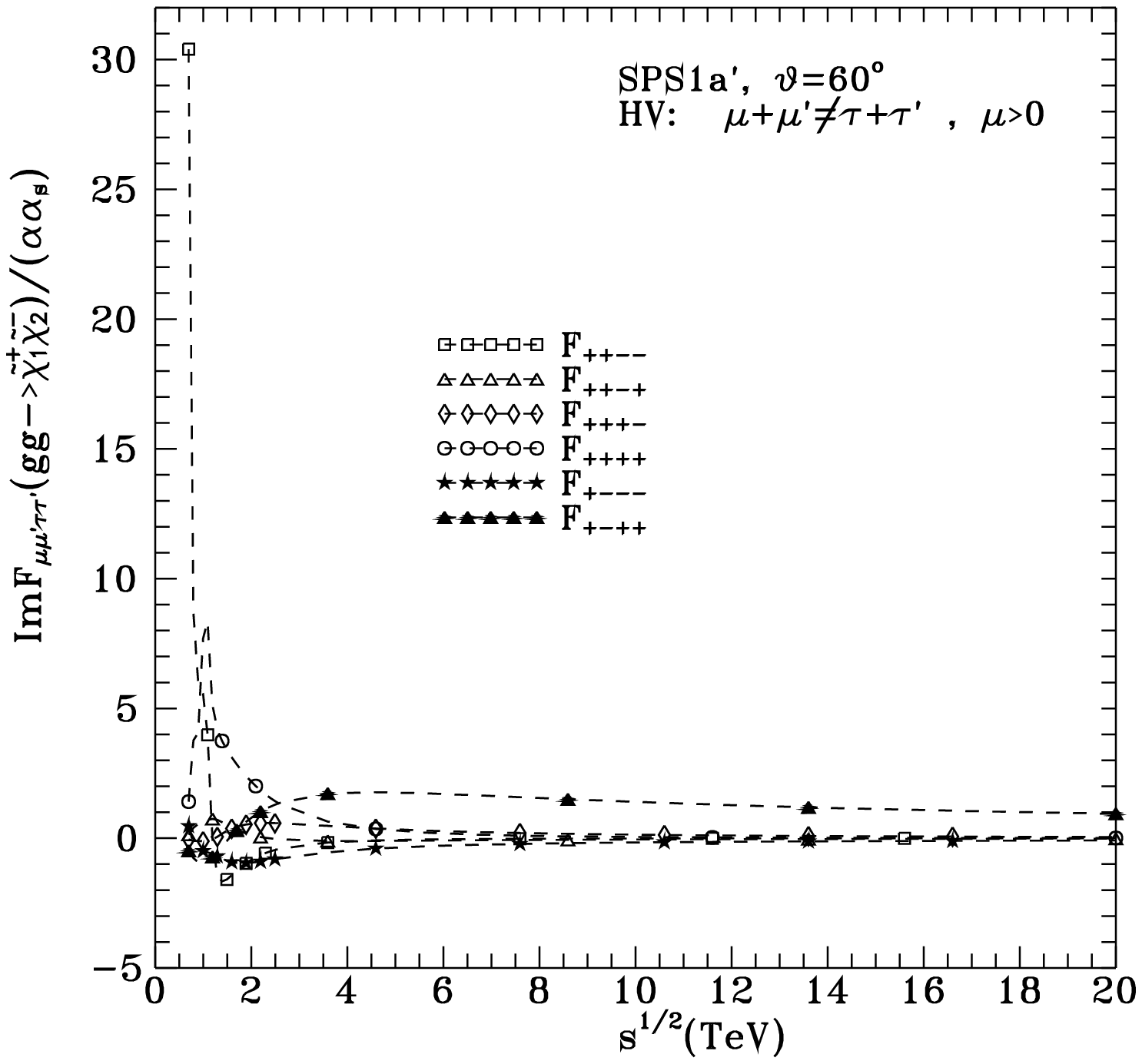,height=6.cm}
\]
\caption[1]{Independent amplitudes for $gg \to \tchi^+_1 \tchi^-_2 $ defined in
 (\ref{Fchipm-ind}), for  $SPS1a'$ at $\theta=60^o$.
The first  row   gives the HC amplitudes, while the  lower ones describe the real and
imaginary parts of the HV amplitudes. Left and right panels
describe respectively amplitudes with negative or  positive helicity $\mu$ for the
first gluon. }
\label{amp-chipm12-SPA-fig}
\end{figure}

We next turn to the exact 1loop EW order amplitudes, for $gg \to \tchi^0_1\tchi^0_2$
and  $gg \to \tchi^+_1\tchi^-_2$, calculated from the ggXXcode  \cite{code} and
given respectively  in Figs.\ref{amp-chi0012-SPA-fig} and
\ref{amp-chipm12-SPA-fig}. Only the independent amplitude
defined  in (\ref{Fchi00-ind}) and (\ref{Fchipm-ind}) are shown, always as functions
of the c.m. energy $\sqrt{s}$, at a fixed c.m. angle $\theta=60^o$.
The upper panels give the HC amplitudes, while the HV amplitudes are shown in  the lower
panels.

As seen in Figs.\ref{amp-chi0012-SPA-fig},\ref{amp-chipm12-SPA-fig}, all HV amplitudes
vanish at high energies rather quickly. Only for  $F_{+-\pm\pm}$ and
$F_{-+\pm\pm}$, the  power-law  vanishing is somewhat delayed by $\ln^2$-terms; compare
(\ref{pmtautau}) and the discussion just before it.

As the energy increases, the HC amplitudes in
Figs.\ref{amp-chi0012-SPA-fig},\ref{amp-chipm12-SPA-fig}, tend to  energy independent,
but angle dependent limits; with the  approach  determined
by powers of the c.m. energy, occasionally delayed  by  logarithmically increasing factors.
For $SPS1a'$ and $\theta=60^o$,
all amplitudes reach their asymptotic
values at around 4 TeV. For other benchmarks, the general shapes
will of course remain the same,
but the energy where asymptopia is reached may  move to higher
or lower values, depending on the SUSY scale
\cite{Baer1, Baer2, Baer3, Baer4, Arnowitt}.\\

\begin{figure}[p]
\vspace*{-1.5cm}
\[
\hspace{-0.5cm}
\epsfig{file=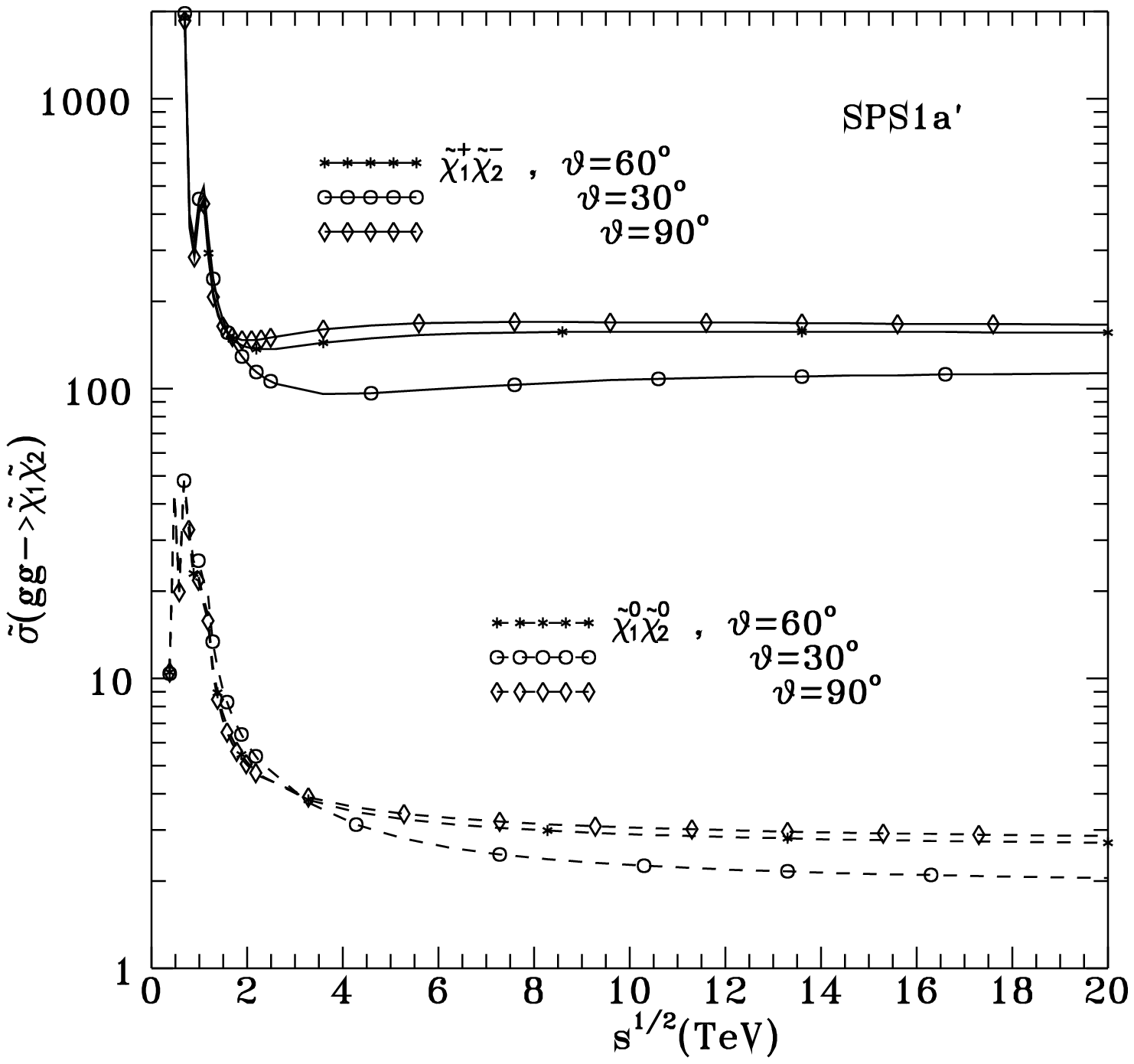, height=7.cm}\hspace{1.cm}
\epsfig{file=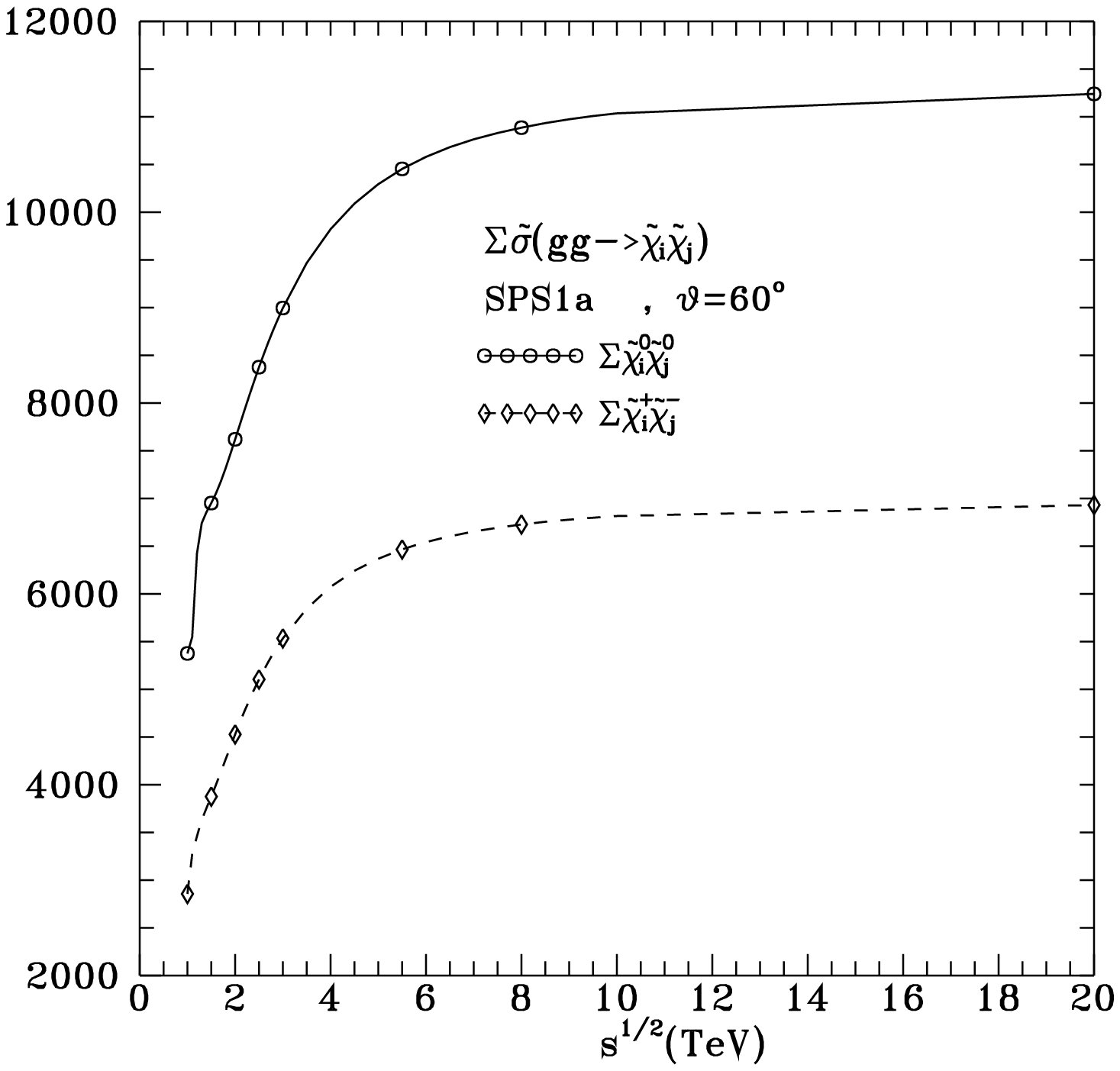, height=7.cm}
\vspace*{0.5cm}
\]
\[
\hspace{-0.5cm}
\epsfig{file=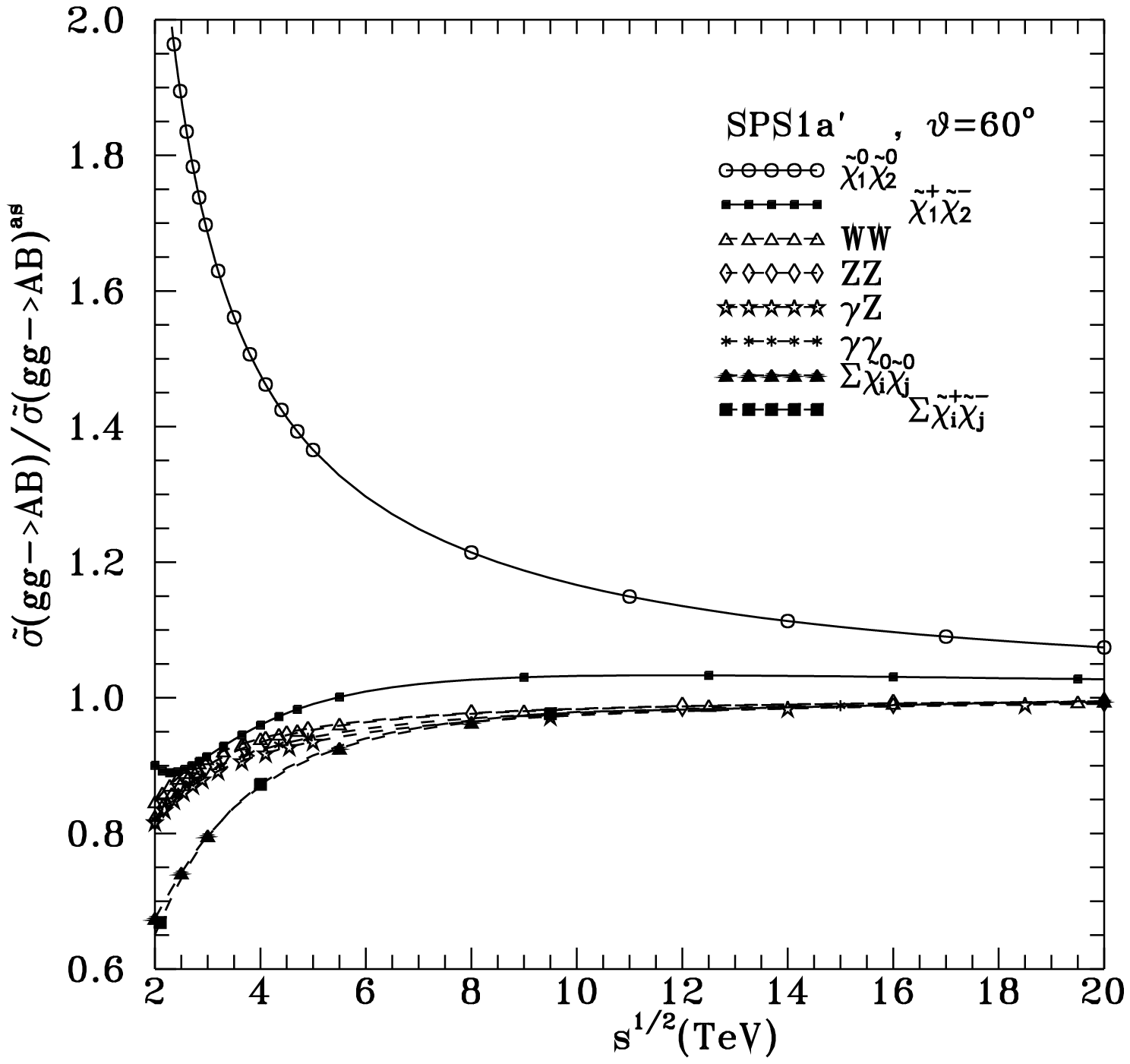, height=7.cm}\hspace{1.cm}
\epsfig{file=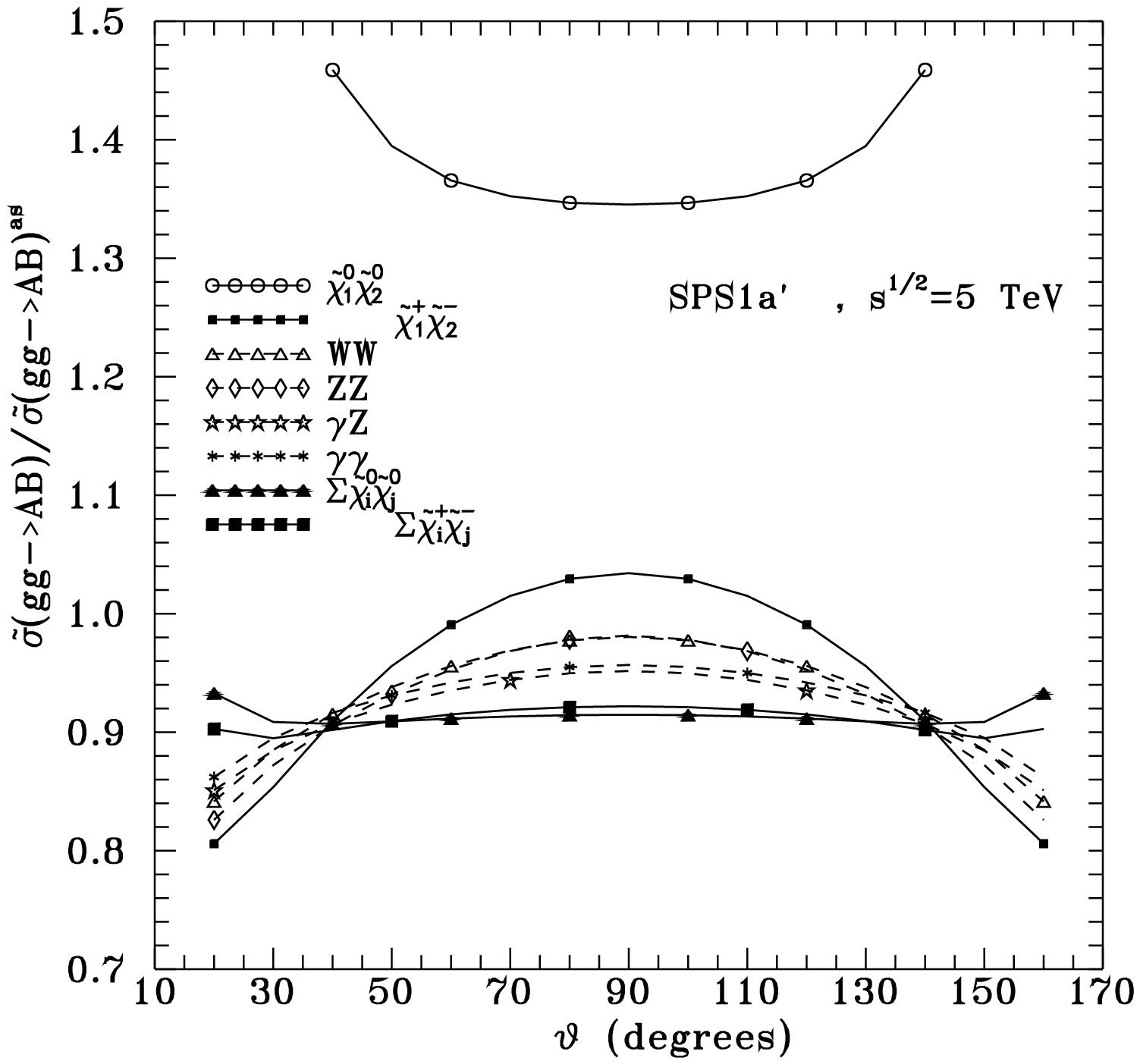, height=7.cm}
\]
\caption[1]{Upper panels: Dimensionless cross sections defined in  (\ref{sigmatilde}),
for $\tilde \sigma(gg \to \tchi^+_1 \tchi^-_2) $ and
 $\tilde \sigma(gg \to \tchi^0_1 \tchi^0_2) $    at $\theta=60^o, ~30^o, ~ 90^o $
(left part);
and  for $\sum_{ij}\tilde \sigma(gg \to \tchi^+_i \tchi^-_j) $,
$\sum_{ij}\tilde \sigma(gg \to \tchi^0_i \tchi^0_j) $ at $\theta=60^0$ (right part).
  Lower panels: Ratios of these dimensionless  cross sections
  to their asymptotic values,   and  the corresponding
  results for $gg \to VV'$, see    (\ref{Rsigma1}, \ref{Rsigma2}, \ref{Rsigma3}).
  Left part displays the energy dependencies
  of these ratios at  $\theta=60^0$, while the right part shows
  the angular dependencies at $\sqrt{s}=5 {\rm TeV}$; compare (\ref{Rsigma-res1}).
  All  in $SPS1a'$. }
\label{sigma-SPA-fig}
\end{figure}

In the left upper panel of   Fig.\ref{sigma-SPA-fig}, we present  the
dimensionless cross sections defined in  (\ref{sigmatilde}),
for $\tilde \sigma(gg \to \tchi^+_1 \tchi^-_2) $ and
 $\tilde \sigma(gg \to \tchi^0_1 \tchi^0_2) $    at $\theta=60^o, ~30^o, ~ 90^o $
 in $SPS1a'$. Correspondingly, in the right upper panel,  we show
 $\sum_{ij}\tilde \sigma(gg \to \tchi^+_i \tchi^-_j) $ and
$\sum_{ij}\tilde \sigma(gg \to \tchi^0_i \tchi^0_j) $ whose asymptotic
benchmark dependence only comes from  $\beta$; in this case only   $\theta=60^0$
is shown.

Unfortunately, for neutralinos,  this summed cross section  is  unobservable,
if  $\tchi^0_1$ is the lightest supersymmetric  particle (LSP), which  contributes
to  Dark Matter.  For charginos though, it may be observable at sufficient energies.
Comparable magnitudes  to these summed cross sections should also always appear
for $\tilde \sigma(gg \to \tchi^0_2 \tchi^0_2) $ and
$\tilde \sigma(gg \to \tchi^+_1 \tchi^-_1) $, which
are of course observable and benchmark dependent.  \\

The lower panels of Fig.\ref{sigma-SPA-fig} intend to show how the various terms in
 (\ref{Rsigma1}, \ref{Rsigma2}, \ref{Rsigma3}) approach 1, as the energy increases.
 As seen there, we find that for $SPS1a'$  and $\sqrt{s} \gsim 5~ {\rm TeV}$,
 \bqa
&&  \frac{\tilde\sigma(gg\to W^+W^-)}{\tilde\sigma(gg\to W^+W^-)^{as}}
 \simeq \frac{\tilde\sigma(gg\to ZZ)}{\tilde\sigma(gg\to ZZ)^{as}}
\simeq  \frac{\tilde\sigma(gg\to \gamma Z)}{\tilde\sigma(gg\to \gamma Z)^{as}}
\simeq  \frac{\tilde\sigma(gg\to \gamma \gamma )}{\tilde\sigma(gg\to \gamma \gamma )^{as}}
  \nonumber \\
&& \simeq \frac{\sum_{ij}\tilde\sigma(gg\to \tchi^+_i\tchi^-_j)}
{\sum_{ij}\tilde\sigma(gg\to \tchi^+_i\tchi^-_j)^{as}} \simeq
  \frac{\sum_{ij}\tilde\sigma(gg\to \tchi_i^0\tchi_j^0)}
{\sum_{ij}\tilde\sigma(gg\to \tchi^0_i \tchi^0_j)^{as}}\simeq 1 ~~, \label{Rsigma-res1}
\eqa
for  a wide range of angles.
Similar pictures  would also hold for all other  benchmarks;
only  the lower bound on energy may   occasionally  need adjustment,
in benchmarks with considerable higher SUSY masses
\cite{Baer1, Baer2, Baer3, Baer4, Arnowitt}.
Apart from the unobservable neutralino member in the last line  of (\ref{Rsigma-res1}),
this relation is an interesting asymptotic SUSY prediction, which may
in fact be valid, even at   LHC type energies. Thus, the gauge and chargino members of
(\ref{Rsigma-res1}), constitute  a prediction, which should  in principle be
observable. In such cases, the experimental data are to be used for the subprocesses
cross sections
in the numerators in (\ref{Rsigma-res1}), while the denominators  are determined by
the simple analytic expressions given in Sect.3.

The situation  in the lower panels of Fig.\ref{sigma-SPA-fig} changes considerably
when specific choices of  $i,j$ are made. Thus for $ \sqrt{s} \simeq  5 ~{\rm TeV}$
in $SPS1a'$ and a wide range of angles,
\bq
 \frac{\tilde\sigma(gg\to \tchi^0_1\tchi^0_2)}{\tilde\sigma(gg\to \tchi^0_1\tchi^0_2)^{as}}
\sim 1.4 ~~; \label{Rsigma-res2}
\eq
 while  $\sqrt{s} \gsim  20 ~{\rm TeV}$ are needed for this quantity to
approach its asymptotic value of 1.
This must be related to the small values of the $Z^N$
matrix elements for $i=1,~j=2$ in $SPS1a'$,  diminishing the contribution of  this channel,
to the complete  dimensionless neutralino  production cross section,
at a  fixed angle; compare  upper panels
of Fig.\ref{sigma-SPA-fig}.

The situation is somewhat better for charginos $\tchi^+_1\tchi^-_2$ in $SPS1a'$;
see lower panel of Fig.\ref{sigma-SPA-fig}, which roughly  satisfies
\bq
 \frac{\tilde\sigma(gg\to \tchi^+_1\tchi^-_2)}{\tilde\sigma(gg\to \tchi^+_1\tchi^-_2)^{as}}
 \sim 1.1 ~~, \label{Rsigma-res3}
\eq
for the same energies and angles. \\

\section{Concluding remarks }

\begin{figure}[p]
\[
\epsfig{file=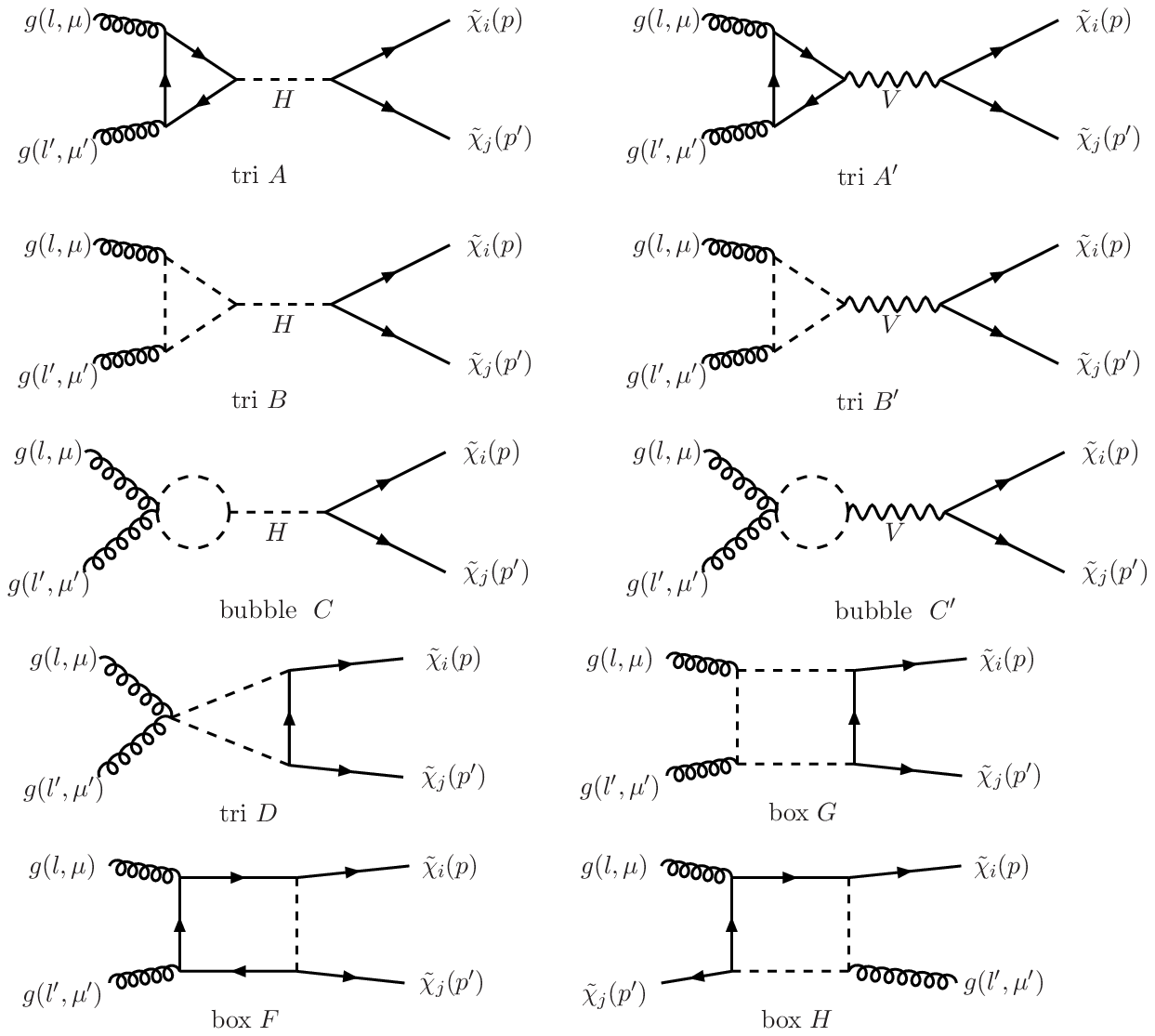,height=11.cm, width=14.cm}
\]
\caption[1]{Independent  graphs for $gg\to \tchi^+_i\tchi^-_j,\tchi^0_i\tchi^0_j$.
Full external lines describe charginos or neutralinos, while full
 internal lines denote  quark or antiquark exchanges. Broken  lines
 describe   squark or antisquark exchanges,
 or the exchange of a neutral MSSM  Higgs-particle denoted as $H$.
 Internal wavy lines describe  neutral electroweak gauge bosons $V$.  }
\label{diagrams-fig}
\end{figure}

In this paper we have analyzed the helicity amplitudes for  $gg\to \tchi_i\tchi_j$
at the 1loop EW level in MSSM, and we have observed the validity of
the Helicity Conservation (HCns) theorem at sufficient energies.
For both, the asymptotically dominant
HC amplitudes, and the suppressed HV ones, the approach to their limiting asymptotic values
is determined by power law expressions, like  $m/\sqrt{s}$ or $m^2/s$,
multiplied by logarithms\footnote{In deriving these results  we focused at the
 sub-sub-leading (i.e. constant) asymptotic contributions the PV expansions \cite{PVasym}.
 This  goes beyond the studies in \cite{MSSMrules1, MSSMrules2, MSSMrules3}
 which only concerned the
leading PV parts.}.
During the calculations, it was fascinating to see how the contributions
of the various diagrams in Fig.\ref{diagrams-fig} were  conspiring in order to assure this.

Very simple expressions for the asymptotic $gg\to \tchi_i\tchi_j$
 HC amplitudes have been  written  in
(\ref{HC-chipmasym}, \ref{HC-chi00asym}), which depend not only on the gauge
couplings and the angle $\beta$,  but also on  ratios of mass-dimension
terms entering the chargino and neutralino mixing matrices. The $m_t/\mw$ and $m_b/\mw$
ratios determining the higgsino contributions, also appear in  (\ref{HC-chi00asym}).

Combining these results, with  the corresponding HC amplitudes  for $gg\to VV'$ derived
in \cite{ggVV} and also appearing in (\ref{HC-VVTTasym}, \ref{HC-VVLLasym}),
the asymptotic subprocess cross section relations in (\ref{Rsigma-res1}) were derived.
Particularly interesting among these, are the relations
 connecting the gauge and total chargino
production through gluon-fusion, in (\ref{Rsigma-res1}). Such relations  may be testable at LHC
type energies, provided   the SUSY scale is in the  TeV range,
as in  \cite{SPA1, SPA2, Baer1, Baer2, Baer3}.
Such relations show that HCns may  have testable implications
at a high energy hadronic collider, even if helicity is not measured directly.
Analogous relations for $ug\to dW^+$ and
$ug\to \tilde d_L \tchi^+_i$were  derived in \cite{ugsdWino}.

 The  ggXXcode released here  may be used
to obtain the corresponding results for any  set of real MSSM parameters
at the EW scale \cite{code}.

Finally, we have also quoted   the  relations between the asymptotic amplitudes for
$gg \to VV'$ and $\tilde g \tilde g \to \tilde V \tilde V', ~ \tilde H_1 \tilde H_1,
~\tilde H_2 \tilde H_2$, appearing  in (\ref{gauge-gaugino-asym}, \ref{Goldstone-higgsino-asym}),
which beautifully illustrate how supersymmetry  manages to preserve
the structure of the amplitudes, in spite of the fact that
the spins of all participating particles are changed.\\

Summarizing, we reiterate that the work in  \cite{heli1, heli2, ugdW, ugsdWino, ggHH, ggVV}
 establishes
the helicity conservation theorem (HCns) for  any 2-to-2 processes,
in the supersymmetric limit of  MSSM with R-parity conservation. This limit
may be  reached by  selecting  the energy to be much larger than all relevant  SUSY masses,
while keeping the scattering angle fixed.

If we try to extend these considerations to  multibody processes,
we immediately realize that complications    increase  rapidly
with the number of  external  particles. Thus,
it is  difficult  to say something general for the analogous
limit, where \underline{all} subenergy squared and momentum transfers become much larger
than the relevant SUSY-masses. Nevertheless,  we may claim
that, if R-parity is conserved, then  all processes  involving an \underline{odd}
number of particles must vanish in this supersymmetric limit  \cite{heli1, heli2}.
Because they will either involve an odd
number of sparticles, or their supersymmetric transformed processes  will do so.

Thus, only processes  involving an even number of particles remain; the least
complicated of which,   are  the  2-to-4 processes.
But even for these, the energy needed for making all   subenergies and momentum
transfers large, becomes prohibitive.
Apparently, it is mainly  for 2-to-2 processes, that
the supersymmetric limit can be studied,  at conceivable energies.

\vspace*{1cm}
\noindent
{\large\bf{Acknowledgements}} GJG is  partially supported by the European Union
 contract MRTN-CT-2006-035505 HEPTOOLS, and the European
 Union ITN programme "UNILHC" PITN-GA-2009-237920.

\clearpage

\clearpage

\begin{thebibliography}{99}


%
\bibitem{heli1}  G.J. Gounaris and F.M. Renard,
\prl{94}{131601}{2005},  hep-ph/0501046.
%
\bibitem{heli2} G.J. Gounaris and F.M. Renard,  \pr{D73}{097301}{2006},  hep-ph/0604041,
(an Addendum).
%
\bibitem{Corfu1} G.J. Gounaris, J. Layssac
and F.M. Renard, \fortp{58}{721}{2010}, e-Print: arXiv:1001.5350 [hep-ph].
%
\bibitem{ugdW} G.J. Gounaris, J. Layssac
and F.M. Renard, \pr{D77}{013003}{2008}, arXiv:0709.1789 [hep-ph].
%
\bibitem{ugsdWino} G.J. Gounaris, J. Layssac
and F.M. Renard, \pr{D77}{093007}{2008}, arXiv:0803.0813 [hep-ph].
%
\bibitem{SPA1} J.A. Aguilar-Saavedra et al., SPA convention,
\epj{C46}{43}{2005}, hep-ph/0511344.
%
\bibitem{SPA2} B.C. Allanach et al. \epj{C25}{113}{2002}, hep-ph/0202233.
%
\bibitem{Baer1} H. Baer, V. Barger, G. Shaughnessy, H. Summy and L-T Wang,
hep-ph/0703289.
%
\bibitem{Baer2} D. Feldman, Z. Liu and P. Nath, \prl{99}{251802}{2007},
arXiv:0707.1873 [hep-ph].
%
\bibitem{Baer3} D. Feldman, Z. Liu and P. Nath, arXiv:0802.4085.
%
\bibitem{ggHH} G.J. Gounaris, J. Layssac and F.M. Renard,
arXiv: 0903.4532 [hep-ph], \pr{D88}{013009}{2009}
%
\bibitem{ggVV} G.J. Gounaris, J. Layssac and F.M. Renard, arXiv:1005.5005,
to appear in the Int. J.  Mod. Phys. A.
%
\bibitem{CM} S. Coleman and J. Mandula, \pr{159}{1251}{1967}.
%
\bibitem{code}  The FORTRAN codes together with a Readme file explaining
its use, are contained in  ggXXcode which can  be downloaded from
http://users.auth.gr/gounaris/FORTRANcodes. All input parameters
in the code are at the electroweak scale.
%
\bibitem{JW} M. Jacob and G.C. Wick, \aop{7}{404}{1959}, \aop{281}{774}{2000}
%
\bibitem{Veltman} G. Passarino and M. Veltman \np{B160}{151}{1979}.
%
\bibitem{looptools1} T. Hahn and M. P\'{e}rez-Victoria, hep-ph/9807565.
%
\bibitem{looptools2} G.J. van Oldenborgh and J.A.M. Vermaseren,
\zp{C46}{425}{1990}.
%
\bibitem{gamgamZZ1}  G.J. Gounaris, J.Layssac, P.I. Porfyriadis
  and F.M.Renard, \epj{C13}{79}{2000}, arXiv:hep-ph/9909243.
  %
\bibitem{gamgamZZ2}    G.J. Gounaris, P.I. Porfyriadis
  and F.M.Renard, \epj{C19}{57}{2001}, arXiv:hep-ph/00100006.
%
\bibitem{PVasym} M. Beccaria, G.J. Gounaris, J. Layssac and F.M. Renard,
\ijmp{A23}{1839}{2008}, arXiv:0711.1067 [hep-ph]  and references therein.
%
\bibitem{couplings} J. Rosiek, \pr{D41}{3464}{1990}.
%
\bibitem{Baer4} H. Baer, T. Krupovnickas, S. Profumo and P. Ullio,
\jhep{510}{020}{2005}, hep-ph/0507282.
 %
\bibitem{Arnowitt} R. Arnowitt aand B Dutta; plenary talk at SUSY02, Hamburg 2002
  hep-ph/0211042.
 %
\bibitem{MSSMrules1}  M. Beccaria,
F.M. Renard and C. Verzegnassi, hep-ph/0203254;
"Logarithmic Fingerprints of Virtual Supersymmetry",
Linear Collider note LC-TH-2002-005,  GDR Supersymmetrie
note GDR-S-081.
%
\bibitem{MSSMrules2}  M. Beccaria, M. Melles, F. M. Renard,
S. Trimarchi, C. Verzegnassi, \ijmp{A18}{5069}{2003}, hep-ph/0304110.
%
\bibitem{MSSMrules3} M. Beccaria,
F.M. Renard and C. Verzegnassi, \ijmp{A24}{623}{2009}, arXiv:0904.2646[hep-ph].
%




\end{thebibliography}
\end{document}